\documentclass[10pt]{iopart}
\makeatletter
\@ifundefined{equation*}{}{%
\expandafter\let\csname equation*\endcsname\@undefined}
\@ifundefined{endequation*}{}{%
\expandafter\let\csname endequation*\endcsname\@undefined}
\makeatother
\usepackage{amsmath, amssymb, amsfonts, amsthm}
\usepackage{mathtools}
\usepackage{bbm}
\usepackage{dsfont}
\usepackage{latexsym}
\usepackage{physics}
\newcommand{\im}{\mathbbm{i}}
\newcommand{\D}{\mathrm{d}}
\usepackage{graphicx}
\graphicspath{{figures/}}
\input{numbers}
\usepackage{booktabs}
\usepackage{multirow}
\usepackage{tabularx}
\usepackage{caption}
\theoremstyle{definition}
\newtheorem{proposition}{Proposition}[section]
\newtheorem{lemma}[proposition]{Lemma}

\newtheorem{definition}{Definition}[section]
\newcommand{\exhead}[2]{\par\medskip\noindent\textbf{Example #1: #2.}\enspace}
\newtheorem{remark}{Remark}[section]
\newtheorem{assumption}{Assumption}

\usepackage{nicefrac}
\usepackage[english]{babel}
\usepackage[T1]{fontenc}
\usepackage[utf8]{inputenc}
\usepackage{footmisc}
\usepackage{iopams}
\usepackage{enumitem}
\usepackage{cite}
\setlist{leftmargin=2em,labelsep=0.6em,itemsep=3pt,topsep=4pt,parsep=0pt,partopsep=0pt}
\usepackage{tikz}
\usetikzlibrary{positioning, fit, arrows.meta, calc}
\begin{document}
\title[]{Subquantum Pilot-Wave System Theory:\\[2pt]
Structural and Amplitude Nonequilibrium under Source Fields}
\author{Said Mikki}
\address{Zhejiang University/University of Illinois at Urbana-Champaign
(ZJU-UIUC) Institute, Zhejiang University, Haining, Zhejiang, China}
\ead{said.m.mikki@gmail.com}
\vspace{10pt}
\begin{abstract}
A source in the Schr\"odinger equation changes the norm of the guiding wave, apparently conflicting with the conservation of probability. In source-extended de~Broglie--Bohm dynamics the conflict disappears: probability is carried by the particle ensemble, while the wavefunction guides its motion. The difficulty is then different. What does quantum equilibrium mean when the particle distribution remains normalized but the guiding density does not? We show that the departure has two independent components: a structural one, measured by the Kullback--Leibler divergence between the particle distribution and the normalized guiding density, and an amplitude one, measured by the Born weight. The Born rule holds precisely when both are equilibrated. Their evolution is governed by the complex source-to-wavefunction ratio; scale covariance further distinguishes an independently prescribed source, against which the amplitude of the wave matters, from state-dependent generators for which rescaling remains a symmetry. We develop a source-aware analytical--computational method, combining exact or semi-analytical guiding fields with Bohmian trajectory transport, ensemble quadrature and independent numerical checks. Free-space impulses expose the limits of global diagnostics and an exact jump rule; a driven ring has constant Born weight but either periodic or secular structural departure; oscillators show both source-generated structural change and amplitude-only evolution, including a finite-horizon weak-drive plateau far below its frozen-particle value; and a coupled two-body model exhibits nonlocal guidance and evolving particle entanglement. The source-free, initially equilibrated sector recovers standard quantum mechanics.

\end{abstract}
\section{Introduction}
\label{sec:introduction}
The de~Broglie--Bohm (dBB) pilot-wave theory is one of the earliest
attempts to build a realist interpretation of quantum mechanics (QM)
by attributing ontological actuality to both the wavefunction $\psi$
and the particle \cite{Bacciagaluppi_Valantini2009}. Developed by de~Broglie in 1927 \cite{deBroglie1927_Solvay} and by Bohm
in 1952 \cite{Bohm1952Part_I,Bohm1952Part_II}, the theory assigns
$\psi$ a double role: as the carrier of the probability wave in
standard QM (SQM), and as the physical field guiding the particle
through configuration space. While the guidance is one-way---the
particle cannot react back on the field---the framework reproduces
all predictions of nonrelativistic QM
\cite{Bohm1952Part_I,Bohm1952Part_II,bohm1993the_undivided_universe,%
durrr2009bohmian_mechanics}, removes the need for an independent
collapse postulate \cite{maudlin2019quantum_philosophy}, and offers
a richer ontological content that has attracted sustained
cross-disciplinary interest
\cite{bell2004speakable,bohm1987quantum_implications,%
Oldofredi2025Guiding_waves_in_quantum_mechanics,%
durrr2009bohmian_mechanics,holland1995the_quantum_theory_of_motion}.
At the same time, the absence of reciprocal interaction and the
ambiguous status of Lorentz and Galilean invariance
\cite{Valentini1997Lorentz,Valentini2025The_trouble} mark the dBB
dynamics as non-mechanistic in a sense that continues to motivate
foundational investigation.

Notably, in Bohm's original 1952 formulation the Born postulate was
clearly separated from the remaining postulates
\cite{Bohm1952Part_I,Bohm1952Part_II}, raising the question of its
fundamental status. Bohm attempted to derive it by introducing a
subquantum level governed by the ratio field $f := P/|\psi|^2$, where
$P$ is the ensemble particle distribution \cite{Bohm1953}; an
alternative stochastic grounding was proposed shortly thereafter
\cite{BohmVigier1954}. A more radical re-evaluation was initiated by
Valentini in the early 1990s
\cite{Valentini1991a,Valentini1991b}, who recognised that
$P = |\psi|^2$ is not a structural necessity of the dBB dynamics but
a contingent ``quantum equilibrium'' state that the theory admits
without enforcing. In this view, $f$ acquires the status of an
independent dynamical variable whose departure from unity signals
\emph{quantum nonequilibrium}. This standpoint contrasts with
the \emph{typicality} programme of D\"urr, Goldstein, and Zangh\`i
\cite{durr1992quantum_equilibrium,durrr2009bohmian_mechanics}, in
which quantum equilibrium is a theorem analogous to Boltzmann's
derivation of thermodynamic equilibrium, and with the
\emph{causal--deterministic} formulation of Holland
\cite{holland1995the_quantum_theory_of_motion}, in which the Born
rule is elevated to an axiom. Against both, Valentini argued that the
nonequilibrium regime $P \neq |\psi|^2$ opens a domain of
\emph{post-quantum} physics---including signal-nonlocal communication
\cite{Valentini1991b,Valentini2025beyond_the_quantum}, violations of
the uncertainty principle \cite{Valentini1991a}, and subquantum
computation \cite{valentini2002subquantum_technology}---accompanied
by a dynamical relaxation mechanism (a subquantum $H$-theorem)
\cite{Valentini1991a,Valentini1991b,valentini2005dynamical,%
towler2011relaxation} and cosmological scenarios in which primordial
nonequilibrium may survive inflation
\cite{valentini2008inflationary,valentini2010inflationary}. The
subquantum dBB theory is therefore best understood as a post-quantum
framework that contains SQM as its equilibrium sector
\cite{Valentini2025beyond_the_quantum,%
valentini2025dBB_Encyclopedia,Valentini2025The_trouble}.

In a previous work \cite{mikki2026AdP}, the
subquantum dBB framework was extended by introducing a complex
Schr\"odinger source field $J(\mathbf{x},t)$ into the wave equation.
Because physical probability resides exclusively in $P$, while $\psi$
acts as a guiding field, the norm of $\psi$ is free to fluctuate
under the influence of $J$ without compromising the conservation of
$P$---a point that dissolves the traditional objection to source
terms in SQM. Borrowing the language of nonequilibrium statistical mechanics
\cite{prigogine2017non-equilibrium}, that work treated the dBB system as
an open system in which $J$ drives it into or out of quantum equilibrium,
while the dynamics itself remains reversible given the history of $J$, and
established the general mathematical structure of the source-extended
dynamics. The framework pointed to a range of phenomena absent from
the source-free scenario---controlled relaxation, source-dependent
regimes of departure from equilibrium, and source-induced modifications of
Bohmian trajectories---whose detailed character was left for further
analysis.

The introduction of $J$, however, also complicates the very notion
of quantum equilibrium. In the source-free theory the condition
$P = |\psi|^2$ is unambiguous: it guarantees Born-rule statistics
\emph{and} agreement with SQM dynamics, because $\psi$ satisfies the
Schr\"odinger equation and $|\psi|^2$ is conserved. With $J \neq 0$
this identification breaks down. The pilot wave, written $\psi_J$ in the
presence of the source, now satisfies an inhomogeneous equation, the Born
density $\rho=|\psi_J|^2$ is generically not conserved, and the velocity field differs from its source-free
counterpart. Born-like position statistics at a single time therefore
no longer entail agreement with SQM dynamics, and the meaning of
``quantum equilibrium'' becomes layered: one must distinguish the
spatial shape of the departure from the overall normalisation of the
guiding wave, and both must be controlled before SQM predictions are
recovered. Moreover, even when Born-like statistics are achieved,
full recovery of SQM requires in addition the absence of the source
itself.

The problem left by introducing a source is therefore not simply whether equilibrium is preserved, but which part of equilibrium can be lost and how that loss appears in the motion of particles. We first separate two questions. Does the particle distribution have the spatial shape of the normalized guiding density? Does the guiding field itself retain unit Born weight? The corresponding diagnostics are the Kullback--Leibler divergence $D$ and the Born weight $N_{\mathrm B}$; the Born rule holds if and only if $D=0$ and $N_{\mathrm B}=1$ (Proposition~\ref{prop:born_two_conditions}). The complex source ratio $J/\psi_J$ governs both rates, but its spatial structure determines how the channels respond differently.

The source also raises a question about amplitude. A rescaling of a source-free guiding field changes no trajectory. With an independently prescribed exogenous source, however, the same rescaling changes the field relative to the source and generally changes the guidance velocity. Proposition~\ref{prop:scale_covariance} makes this distinction precise; state-dependent linear generators retain the usual covariance. Whether absolute field and source amplitudes are themselves physical is part of the ontology adopted here, not a theorem inferred from the trajectories.

The analytical classification does not yet tell us what driven ensembles do. For that we solve the sourced field equation by Duhamel integration and follow the particle distribution along Bohmian characteristics. Exact or semi-analytical field solutions are combined with numerical characteristic integration and ensemble quadrature where closed forms end, and are tested by independent checks (\ref{app:numerics}). This analytical--computational investigation is a second contribution of the paper: it reveals regimes that the equilibrium definitions alone do not predict.

The examples in Sec.~\ref{sec:examples} develop this investigation progressively. Infinite-space impulses identify what can be said about local guidance when global normalized diagnostics fail; a regularized pulse yields an exact jump rule. The ring then shows that a fixed Born weight can coexist with periodic or secular structural evolution. Driven harmonic oscillators distinguish source-generated structural departure, amplitude-only dynamics, coherent-state response and the effect of changing initial field amplitude against a fixed source. A coupled two-body oscillator finally connects the same diagnostics with nonlocal guidance and entanglement. The weak-drive oscillator results concern long but finite numerical windows, not a proven asymptotic limit. Standard quantum mechanics is recovered when $J=0$ and the initial state is in full quantum equilibrium, $D(0)=0$ and $N_{\mathrm B}(0)=1$; source-free nonequilibrium instead preserves its initial diagnostics.

Section~\ref{sec:general-structure} develops the $N$-body framework and equilibrium classification. Section~\ref{sec:formalism} gives the field and particle solution theory and distinguishes source types; Section~\ref{sec:examples} investigates the examples; and Section~\ref{sec:conclusion} draws the conclusions. The mathematical prerequisites, proof of scale covariance, numerical verification and conditional operational reconstruction are collected in \ref{app:math}, \ref{app:scale_proof}, \ref{app:numerics} and \ref{app:measure}.

\section{General Structure of the Subquantum System}
\label{sec:general-structure}
The distinction between wave and particle probability is clearest in configuration space, where both the guiding field and the particle ensemble are defined. Let $\mathcal C=\mathbb R^{3N}$ for $N$ particles, with positions $\mathbf x_i\in\mathbb R^3$ and velocities $\mathbf v_i\in\mathbb R^3$. Write $x=(\mathbf x_1,\ldots,\mathbf x_N)^\top$ and $v=(\mathbf v_1,\ldots,\mathbf v_N)^\top$. The de~Broglie field is a complex scalar $\psi:\mathcal C\to\mathbb C$; its guidance velocity is denoted $v^\psi$ to make the dependence on the field explicit.

A trajectory $q(t;x_0)$ starts at $x_0\in\mathcal C$ and collects the individual paths $\mathbf q_i(t)$ in configuration space. The velocity $v^\psi(x,t)$ is the Eulerian field, while $v^\psi(q(t;x_0),t)=\mathrm dq/\mathrm dt$ is its Lagrangian value along a path.\footnote{In fluid mechanics an Eulerian quantity is specified at fixed position and time, while a Lagrangian quantity is evaluated along a moving particle.} This distinction will matter when we compute diagnostics by averaging over initial particle positions rather than over a reconstructed density at every later time.

\subsection{General View and Ontological Framework}
\label{subsec:ontological-framework}
The extended theory contains three fields on $\mathcal C$: the pilot wave $\psi_J:\mathcal C\times\mathbb R\to\mathbb C$, the particle probability distribution $P:\mathcal C\times\mathbb R\to[0,\infty)$, and the source $J:\mathcal C\times\mathbb R\to\mathbb C$. All live in configuration space, but $P$ is not identified ontologically with $|\psi_J|^2$. The theory is therefore an ensemble theory in the sense relevant to information-geometric formulations (see, e.g., \cite{HallReginatto2016Ensembles}). Like ordinary dBB theory it uses an absolute time $t\in\mathbb R$, consistent with its nonlocal ontology \cite{valentini2025dBB_Encyclopedia}.

\begin{figure}[htbp]
\centering
\begin{tikzpicture}[
node distance=2.5cm and 3.5cm,
block/.style={rectangle, draw, fill=blue!5, text width=5.5cm,
minimum height=3cm, align=center, rounded corners,
font=\small},
inputblock/.style={rectangle, draw, fill=green!5, text width=2.5cm,
minimum height=1.5cm, align=center,
rounded corners, font=\small},
outputblock/.style={rectangle, draw, fill=orange!5,
text width=2.5cm, minimum height=1.5cm,
align=center, rounded corners, font=\small},
arrow/.style={-Stealth, thick, color=black!80},
lbl/.style={font=\small\itshape, color=black!70}
]
\node[inputblock] (init_state)
{Initial State \\ $\Sigma_{t_0} = (\psi_0, P_0)$};
\node[inputblock, below=1.2cm of init_state] (source)
{Source Field \\ $J(x, t)$ for $t \ge t_0$};
\node[block, right=3.5cm of init_state] (system) {
\textbf{Subquantum Dynamics} \\[0.2cm]
\textit{Field Sector:} \\
$\im\hbar \partial_t \psi_J = H \psi_J + J$ \\[0.2cm]
\textit{Particle Sector:} \\
$\dot{x} = v^{\psi_J}(x, t)$ \\
$\partial_t P + \nabla \cdot (P v^{\psi_J}) = 0$
};
\node[outputblock, below=2.2cm of system] (output_state)
{Evolved State \\ $\Sigma_t = (\psi_J(\cdot, t), P(\cdot, t))$};

\draw[arrow] (init_state.east) --
node[lbl, above] {Initial Condition}
(system.west |- init_state.east);
\draw[arrow] (source.east) -|
node[lbl, pos=0.25, below] {Continuous Driving}
([xshift=-1.2cm]system.south);
\draw[arrow] (system.south) --
node[lbl, right] {State at $t$}
(output_state.north);

\node[draw, dashed, thick, fit=(system) (output_state),
inner sep=12pt,
label=above:\textbf{Non-Autonomous Evolution Operator
$\mathcal{E}(t, t_0)$}] {};
\end{tikzpicture}
\caption{Block diagram of the extended subquantum dynamics: the initial state and the source field drive the field and particle sectors, and the non-autonomous evolution operator $\mathcal{E}(t,t_0)$ produces the evolved state $\Sigma_t$.}
\label{fig:subquantum_system}
\end{figure}
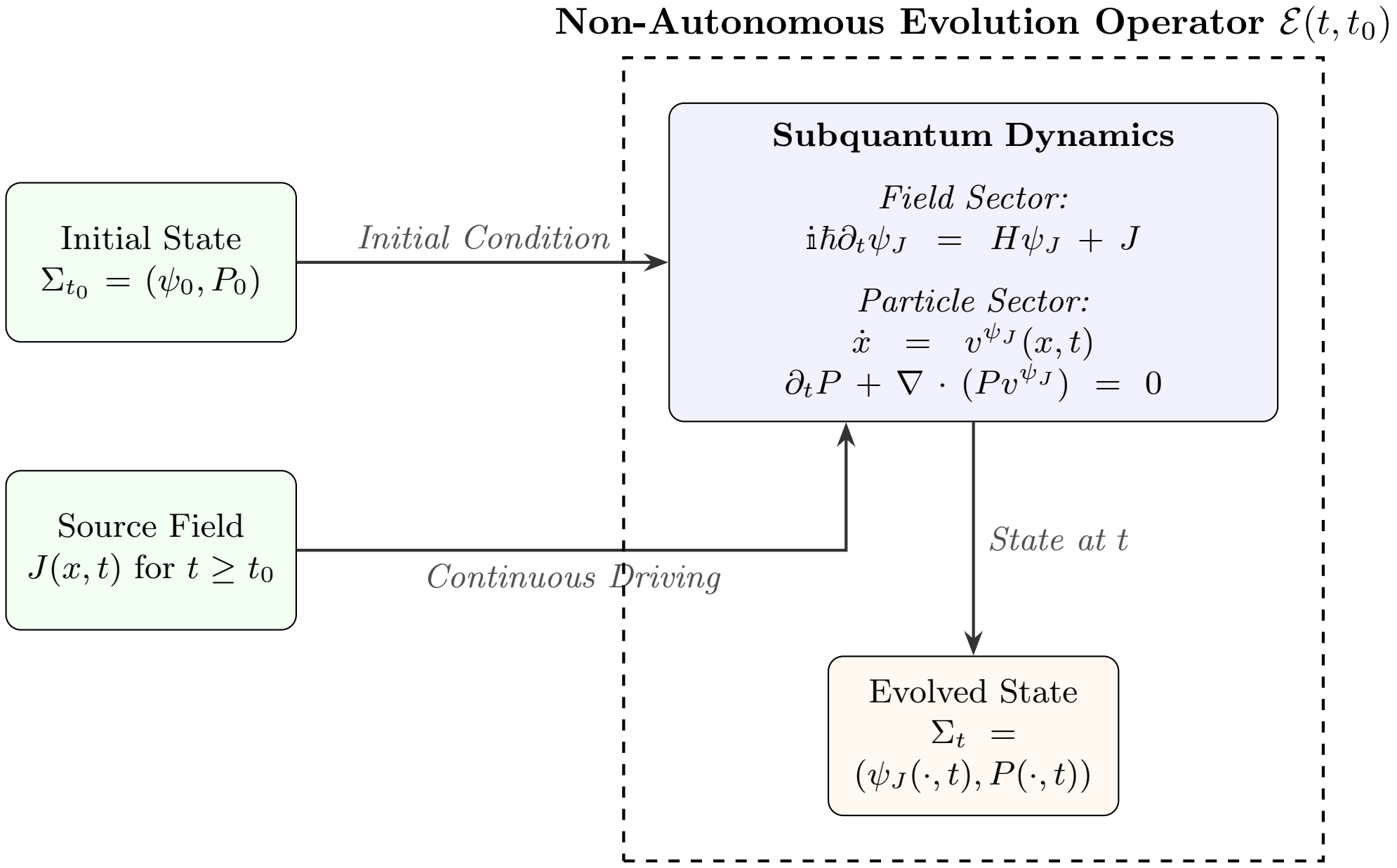

Its causal structure has two layers. The field sector evolves $\psi_J$ under the inhomogeneous Schr\"odinger equation (ISE)

\begin{equation}
\im\hbar \frac{\partial \psi_J}{\partial t} = H \psi_J + J,
\label{eq:ise}
\end{equation}
Here $H$ is self-adjoint on a dense domain of $\mathcal H=L^2(\mathcal C)$, and $J$ is a complex source. The complete ontological state is $\Sigma_t=(\psi_J(\cdot,t),P(\cdot,t))\in\mathcal H\times\mathfrak P(\mathcal C)$, with $\mathfrak P(\mathcal C)$ the space of probability measures. An admissible input is $J\in L^1_{\mathrm{loc}}([0,\infty);\mathcal H)$,\footnote{Thus $\int_0^T\|J(\cdot,t)\|_{L^2}\,\mathrm dt<\infty$ for every finite $T$; function-space conventions are stated in \ref{app:math}.} and the input and initial data determine $\Sigma_t$ (Fig.~\ref{fig:subquantum_system}). Unlike source-free evolution $e^{-\im Ht/\hbar}\psi_0$, the solution depends on the history of $J$, as the Duhamel representation in Sec.~\ref{sec:duhamel} will show.

The norm need not be conserved: $\mathrm d\|\psi_J\|_{L^2}^2/\mathrm dt=(2/\hbar)\operatorname{Im}\langle\psi_J,J\rangle_{\mathcal H}$. The source can inject or extract guiding-field amplitude without introducing an intrinsic arrow of time. This is not Lindblad-type dissipation, whose dynamics is irreversible and trace-distance contractive \cite{Lindblad1976,GoriniKossakowskiSudarshan1976}: with the source history specified, the ISE remains affine and reversible, with no intrinsic attractor.

The field alone does not describe the particle ensemble. In the particle sector each configuration follows the Bohmian guidance velocity

\begin{equation}
\mathbf{v}_i^{\psi_J} = \frac{1}{m_i} \nabla_{\mathbf{x}_i} S_J, \qquad i \in \{1, \dots, N\},
\label{eq:guidance}
\end{equation}
where $\psi_J = R_J e^{\im S_J/\hbar}$ is the polar decomposition of the pilot wave, and the distribution is advected by the same field through the continuity equation
\begin{equation}
\frac{\partial P}{\partial t}
+ \sum_{i=1}^N \nabla_{\mathbf{x}_i}
\cdot \left( P \mathbf{v}_i^{\psi_J} \right) = 0,
\label{eq:continuity}
\end{equation}
The continuity equation transports $P_0$ into $P_t$ and defines the ensemble evolution, with individual trajectories serving as its characteristics; different trajectory descriptions can correspond to the same ensemble density \cite{deotto1998bohmian}. Whereas standard QM describes the state by a ray $[\psi]\in\mathcal H$ \cite{Sontz2020Quantum_probability}, the extended ontology requires the pair $(\psi_J,P)$. Its causal order is

\begin{equation}
J \;\longrightarrow\; \psi_J \;\longrightarrow\; v^{\psi_J}
\;\longrightarrow\; P,
\label{eq:causal_hierarchy}
\end{equation}
The arrows have a one-way meaning. The source drives $\psi_J$ through Eq.~\eqref{eq:ise}; the guiding field determines $v^{\psi_J}$ through Eq.~\eqref{eq:guidance}; and the velocity advects $P$ through Eq.~\eqref{eq:continuity}. Neither the particles nor their probability distribution react back on $\psi_J$, and $\psi_J$ does not determine the independently prescribed $J$. Consequently the state is determined by the source history and $(\psi_0,P_0)$.

A source may even generate a guiding wave from a null initial field, $\psi_J(x,t_0)=0$ with $P_0\ne0$. This removes the need to posit an earlier guiding wave as an initial condition, but leaves the origin of the source itself outside the theory. The singular initial instant requires the separate treatment given in Remark~\ref{rem:null_start}.

\subsection{$N$-Body Configuration-Space Formulation}
\label{sec:nbody}
The preceding structure extends directly to $N$ particles. We state its configuration-space form, with $\mathcal C=\mathbb R^{3N}$, before restricting the examples to reduced one- and two-dimensional models. Let $x=(\mathbf x_1,\ldots,\mathbf x_N)$, with masses $m_i$, and use $\mathrm d\Omega=\mathrm d^3x_1\cdots\mathrm d^3x_N$. The derivation parallels the one-body equations of Ref.~\cite{mikki2026AdP}. The Hamiltonian is $H=-\sum_{i=1}^N(\hbar^2/2m_i)\nabla_i^2+V$, and for an exogenous source $J$ is independently prescribed rather than a functional of $\psi_J$ (Definition~\ref{def:source_types}). With $\psi_J=R_Je^{\im S_J/\hbar}$, the guidance law is $\dot{\mathbf q}_i=\mathbf v_i^{\psi_J}$, which also has the complex-ratio form

\begin{equation}
\mathbf{v}^{\psi_J}_i
=
\frac{\hbar}{m_i}
\operatorname{Im}\!\left(
\frac{\nabla_i \psi_J}{\psi_J}
\right).
\label{eq:nbody_v_complex}
\end{equation}
For $N \geq 2$, the velocity of particle $i$ \emph{can} depend on the positions of other particles through a nonseparable configuration-space wavefunction $\psi_J$, which is the origin of nonlocal guidance in dBB theory. The source enters through the \emph{complex source ratio}, which unifies the phase-source and Born-density source terms into a single quantity,
\begin{equation}
W(x,t) := \frac{J(x,t)}{\psi_J(x,t)}
= \operatorname{Re} W(x,t) + \frac{\im\hbar}{2}\Gamma(x,t),
\label{eq:W_definition}
\end{equation}
defined pointwise where $\psi_J \neq 0$, with
\begin{equation}
\operatorname{Re} W
=
\frac{1}{R_J}\operatorname{Re}\!\left[
J e^{-\im S_J/\hbar} \right],
\qquad
\Gamma
:=
\frac{2}{\hbar}\operatorname{Im}\!\left(\frac{J}{\psi_J}\right)
= \frac{2}{\hbar}\operatorname{Im} W.
\label{eq:W_components}
\end{equation}
The real part $\operatorname{Re} W$ enters the modified Hamilton--Jacobi equation~\eqref{eq:nbody_hj}, while the imaginary part $\operatorname{Im} W = \hbar\Gamma/2$ determines the source term of the Born-density balance equation~\eqref{eq:nbody_continuity} and hence governs the amplitude channel.

The formulas for $\mathbf v_i^{\psi_J}$ and, when $J\ne0$, for $\Gamma$ are undefined at nodes of $\psi_J$. Near them, trajectories can move rapidly and the source rate can be singular. Local solutions of the guidance equation may also fail to exist for an entire interval because a trajectory can approach a node or escape to infinity. The integral identities that follow therefore require explicit hypotheses; they must not be read as a general global-existence theorem.

We use three standing assumptions on an arbitrary finite interval $[0,T]$. Here $P_0$-almost everywhere (a.e.) means except on a set of zero initial ensemble measure.\footnote{A set of starting positions of $P_0$-measure zero contributes no probability to the initial ensemble.} Assumption~A gives a unique local path on the nonnodal domain. Assumption~B separately requires almost every such path to survive up to $T$. Assumption~C controls the rate integrals, keeps the Born weight positive and finite whenever normalized diagnostics are used, and excludes leakage through spatial infinity (or accounts for periodic flux cancellation on the ring). Together they provide the conditions under which conservation and normalization can be used in the source-driven theory.

\begin{assumption}[Standing regularity]
\label{ass:A}
On the nonnodal spacetime domain $\mathcal{U}=\{(x,t):\psi_J(x,t)\neq0\}$ the Bohmian velocity field $v^{\psi_J}$ is continuous in time and locally Lipschitz in space, and $P_0(\mathcal{G}_0)=1$, so that the guidance ODE~\eqref{eq:guidance} has a unique maximal local solution $q(t;x_0)$ for $P_0$-a.e.\ initial condition $x_0$. Local existence and uniqueness do not by themselves give completeness, which is the content of Assumption~\ref{ass:H}. It holds, for instance, when $\psi_0$ and $J(\cdot,t)$ lie in the domain of a sufficiently high power of $H$ and $J$ is smooth in time (\ref{rem:duhamel_wellposed}).
\end{assumption}

\begin{assumption}[Trajectory completeness]
\label{ass:H}
$P_0\bigl(\{x_0:\ [0,T]\not\subset I(x_0)\}\bigr)=0$, where $I(x_0)$ is the maximal existence interval of $q(\cdot;x_0)$; that is, $P_0$-almost every trajectory exists on the whole closed interval $[0,T]$, including the endpoint. This is an assumption on the initial ensemble, not a consequence of the local regularity condition, and its verification for particular sources is a separate matter.
\end{assumption}

\begin{assumption}[Finiteness of rates]
\label{ass:E}
On $[0,T]$: (i) $\int_0^T\!\bigl(\langle|\Gamma|\rangle_P + \langle|\Gamma|\rangle_\rho\bigr)\,\mathrm{d}t < \infty$, where $\langle g\rangle_P=\int g\,P\,\mathrm{d}\Omega$ and $\langle g\rangle_\rho=N_{\mathrm B}^{-1}\int g\,\rho\,\mathrm{d}\Omega$; (ii) $0<N_{\mathrm{B}}(t) < \infty$ throughout the interval on which the normalized diagnostics are used; and (iii) the integrated flux contributions of $\rho v$ and $Pv$ through the cutoff surfaces vanish, that is, $\int_0^t\!\int(\rho v)\cdot\nabla\chi_R\,\mathrm{d}\Omega\,\mathrm{d}s\to0$ and $\int_0^t\!\int(Pv)\cdot\nabla\chi_R\,\mathrm{d}\Omega\,\mathrm{d}s\to0$ as $R\to\infty$ for $t\le T$ and smooth cutoffs $\chi_R\to1$ on the Cartesian configuration space $\mathbb{R}^{d}$ ($d=3N$ in general and $d=1,2$ in the reduced examples), and these contributions cancel by periodicity in the ring model.
\end{assumption}

These conditions govern the continuum derivations; they are not a blanket claim that all illustrative sources satisfy them. Normalization and rate conditions can be checked directly for the ring (Sec.~\ref{sec:ring}) and the amplitude-only oscillator HO.2 (Sec.~\ref{sec:ho2}). In the other oscillator and two-body models, $\Gamma$ may be unbounded near field nodes and its integrability along the full ensemble remains conditional. The finite-horizon trajectories reported in Sec.~\ref{sec:examples} support the calculations but do not establish almost-everywhere completeness (\ref{app:numerics}). Ideal impulses and null starts are handled piecewise or formally: $\psi_0=0$ with $P_0>0$ does not satisfy $P_0(\mathcal G_0)=1$ at the initial instant.

Substituting the polar decomposition
into~\eqref{eq:ise} yields the modified Hamilton--Jacobi equation
\begin{equation}
\frac{\partial S_J}{\partial t}
+
\sum_{i=1}^N \frac{(\nabla_i S_J)^2}{2m_i}
+ V
+ Q^{\mathrm{B}}_J
=
-\operatorname{Re} W,
\label{eq:nbody_hj}
\end{equation}
where the Bohm potential on $\mathcal{C}$ is $Q^{\mathrm{B}}_J:=-\sum_{i=1}^N(\hbar^2/2m_i)\nabla_i^2R_J/R_J$, and the modified continuity equation for the Born density
$\rho := |\psi_J|^2 = R_J^2$ reads
\begin{equation}
\frac{\partial \rho}{\partial t}
+
\sum_{i=1}^N \nabla_i \cdot
(\rho\,\mathbf{v}^{\psi_J}_i)
=
\Gamma\,\rho.
\label{eq:nbody_continuity}
\end{equation}
The continuity equation~\eqref{eq:continuity} for the particle
distribution $P$ remains source-free. (Here and throughout, the
conservation $\int P\,\mathrm{d}\Omega = 1$ holds under
Assumption~\ref{ass:H}; without it, mass can be lost at terminating
trajectories.)

Working with the Bohm--Valentini ratio field $f := P/\rho$
\cite{Bohm1953,Valentini1991a}, the modified transport equation on
$\mathcal{C}$ is
\begin{equation}
\frac{\mathrm{D} f}{\mathrm{D} t}
:=
\frac{\partial f}{\partial t}
+
\sum_{i=1}^N \mathbf{v}^{\psi_J}_i \cdot \nabla_i f
=
-\Gamma f,
\label{eq:nbody_f_transport}
\end{equation}
where $\mathrm{D}/\mathrm{D}t$ is the material derivative following a Bohmian particle, as in hydrodynamic transport theory, and $\Gamma$, of dimension inverse time, is the rate at which the source modifies $f$ along a trajectory. The negative sign arises because the particle density is source-free whereas the Born density obeys Eq.~\eqref{eq:nbody_continuity}.

Along a configuration-space trajectory
$q(t) = (\mathbf{q}_1(t),\ldots,\mathbf{q}_N(t))^\top$ with initial
configuration $q(0) = x_0$, we define the trajectory-restricted ratio
field
\begin{equation}
F(t;x_0) := f\bigl(q(t;x_0), t\bigr),
\label{eq:nbody_F_definition}
\end{equation}
so that the transport equation~\eqref{eq:nbody_f_transport} reduces to
the ordinary differential equation
\begin{equation}
\frac{\mathrm{d} F}{\mathrm{d} t}(t;x_0)
=
-\Gamma(t)\,F(t;x_0),
\label{eq:nbody_F_ode}
\end{equation}
where $\Gamma(t) := \Gamma(q(t;x_0), t)$ is evaluated along the same
trajectory. Here $F$ is the restriction of $f$ to a characteristic of
Eq.~\eqref{eq:nbody_f_transport}, which is a Bohmian trajectory.

\subsection{Normalized Diagnostics and the Entropy Decomposition}
\label{sec:diagnostics}
The source-free theory admits a single equilibrium condition: $P=\rho=|\psi|^2$, equivalently $f=P/\rho=1$. The two densities obey the same continuity equation, and the Born norm is conserved. With $J\ne0$ the comparison changes. The particle distribution is still normalized under Assumption~\ref{ass:H}, but the Born density has the additional term $\Gamma\rho$ in Eq.~\eqref{eq:nbody_continuity}. Its norm can vary even when its spatial profile continues to agree with $P$. We must therefore separate the normalization of the guiding field from its shape.

The Born weight on $\mathcal{C}$ is
\begin{equation}
N_{\mathrm{B}}(t)
:=
\int_{\mathcal{C}} \mathrm{d}\Omega\; \rho(x,t).
\label{eq:nbody_NB}
\end{equation}
\begin{remark}[Dimensional conventions]
Throughout, the theory remains dimensional, with explicit $\hbar$, masses and frequencies, and we adopt the usual probability-density convention in the physical Cartesian coordinates of $\mathcal{C}=\mathbb{R}^{d}$, with $d=3N$ in general and $d=1,2$ in the reduced examples. Since only $\int P\,\mathrm{d}\Omega=1$ is a pure number, $P$ is a density of dimension $L^{-d}$, and for $P=|\psi_J|^2$ to hold at full equilibrium the guiding field and the source must carry $[\rho]=[P]=L^{-d}$, $[\psi_J]=L^{-d/2}$ and $[J]=E\,L^{-d/2}$, where $E$ denotes energy. The Born weight $N_{\mathrm B}$, the ratios $f$, $F$ and $\tilde f$, and the diagnostic $D$ are then dimensionless, so that the condition $N_{\mathrm B}=1$ requires no reference norm and every logarithm in this paper has a dimensionless argument, while $[W]=E$ and $[\Gamma]=T^{-1}$ follow from the ISE~\eqref{eq:ise}, which fixes only the relative units $[J]/[\psi_J]=E$.

Two distinctions prevent misreadings. First, choosing units is not the physically distinct operation of rescaling $\psi_J$ at fixed exogenous $J$ (Sec.~\ref{subsec:source_types}): the latter changes $J/\psi_J$ and in general the dynamics, whereas a change of the amplitude unit changes both fields consistently.

Second, the dimension of a source \emph{coefficient} depends on how its profiles are normalized and should not be identified with $[J]$: for a modal source $J=j_0e^{-\im\Omega t}\varphi_n(x)$ with a normalized eigenfunction $\varphi_n$ of the same density-amplitude dimension as $\psi_J$, the coefficient $j_0$ is an energy, while for a pulse $J=j_0\,g(x)\,h_\varepsilon(t)$ with $[g]=[\psi_J]$ and $\int h_\varepsilon\,\mathrm{d}t=1$ it is an energy times a time. On the ring the integration measure is the arclength. The numerical examples use declared scaled units, which are specified at the beginning of Sec.~\ref{sec:examples} and in \ref{app:numerics}.
\end{remark}

To derive its evolution, integrate the modified continuity
equation~\eqref{eq:nbody_continuity} over $\mathcal{C}$. The divergence
term $\sum_{i=1}^N \nabla_i \cdot
(\rho\,\mathbf{v}_i^{\psi_J})$ vanishes upon integration (by the no-leakage condition of Assumption~\ref{ass:E}, that is, sufficient decay at infinity on $\mathbb{R}^{3N}$ or periodic boundary conditions on the ring), leaving
\begin{equation}
\frac{\mathrm{d} N_{\mathrm{B}}}{\mathrm{d} t}
=
\int_{\mathcal{C}} \mathrm{d}\Omega\; \Gamma(x,t)\,\rho(x,t)
=
\langle \Gamma \rangle_\rho \, N_{\mathrm{B}}(t),
\label{eq:nbody_NB_rate}
\end{equation}
where the $\rho$-weighted average of $\Gamma$ is $\langle \Gamma \rangle_\rho:=N_{\mathrm{B}}^{-1}\int_{\mathcal{C}}\mathrm{d}\Omega\,\Gamma\rho$ and the $P$-weighted average is defined likewise, $\langle\Gamma\rangle_P := \int_{\mathcal{C}}\mathrm{d}\Omega\,\Gamma P$.
Equation~\eqref{eq:nbody_NB_rate} shows that $\Gamma(x,t)$ is the local
rate at which Born density is created ($\Gamma>0$) or destroyed
($\Gamma<0$) per unit of $\rho$: the sign of $\langle\Gamma\rangle_\rho$
decides whether the Born weight grows or decays, while the spatial
fluctuations of $\Gamma$ decide where the production or destruction is
concentrated. By contrast, $\int_{\mathcal{C}}\mathrm{d}\Omega\,P = 1$ for
all $t$ regardless of $J$, since Eq.~\eqref{eq:continuity} is
source-free (under Assumption~\ref{ass:H}). In particular, $N_{\mathrm{B}}$
is conserved on $[0,T]$ if and only if $\langle\Gamma\rangle_\rho \equiv 0$
there, and $f \equiv 1$ cannot persist unless $\Gamma \equiv 0$ on
$\mathcal{G}_t := \{x : \psi_J(x,t) \neq 0\}$ (equivalently, $\rho > 0$),
the set on which $\Gamma$ and $f$ are defined; this is why we work with
the Born weight.

\begin{remark}[Square-Integrability of the Born Weight]
\label{rem:NB_square_integrability}
Equation~\eqref{eq:nbody_NB_rate} requires $N_{\mathrm{B}} < \infty$, which
fails for idealized sources that do not decay at infinity: plane waves,
point sources, or sources depending only on relative coordinates and
hence constant along the centre-of-mass direction. The corresponding
solutions are not square-integrable and the normalized quantities below
are undefined. Such infinite-space examples
(Sec.~\ref{sec:free_infinite}) are therefore used only for local
kinematics, whereas the full diagnostics are applied to the finite-volume
and localized examples (Secs.~\ref{sec:ring}--\ref{sec:twobody}).
\end{remark}

To restore a well-defined equilibrium reference in the presence of a
time-varying Born weight, we introduce the normalized Born density
\begin{equation}
\tilde{\rho}(x,t) := \frac{\rho(x,t)}{N_{\mathrm{B}}(t)},
\label{eq:nbody_rho_tilde}
\end{equation}
which satisfies
$\int_{\mathcal{C}} \mathrm{d}\Omega\; \tilde{\rho} = 1$
by construction, and the normalized Bohm--Valentini ratio field
\begin{equation}
\tilde{f}(x,t)
:=
\frac{P(x,t)}{\tilde{\rho}(x,t)}
=
N_{\mathrm{B}}(t) \, f(x,t).
\label{eq:nbody_ftilde}
\end{equation}
The subquantum entropy functional of Valentini \cite{Valentini1991a},
expressed on the configuration space $\mathcal{C}$, is
\begin{equation}
S_{\mathrm{e}}(t)
:=
-k_{\mathrm{B}} \int_{\mathcal{C}} \mathrm{d}\Omega \;
P(x,t) \ln f(x,t),
\label{eq:nbody_entropy_unnormalized}
\end{equation}
where $k_{\mathrm{B}}$ is Boltzmann's constant; for $J = 0$ it equals
$-k_{\mathrm{B}}$ times the Kullback--Leibler divergence of $P$ from the
normalized density $\rho$. With a source the Born density is no longer
normalized, and $S_{\mathrm{e}}$ mixes two contributions: the
redistribution of $P$ relative to the \emph{shape} of $\rho$ (structural
channel) and a global shift of magnitude
$k_{\mathrm{B}} \ln N_{\mathrm{B}}(t)$ produced by the rescaling of the
Born norm (amplitude channel). To treat them separately we use the
normalized quantities $\tilde{\rho}$ and $\tilde{f}$ of
Eqs.~\eqref{eq:nbody_rho_tilde}--\eqref{eq:nbody_ftilde} and define the
structural diagnostic
\begin{equation}
D(t)
:=
\int_{\mathcal{C}} \mathrm{d}\Omega \;
P \ln \frac{P}{\tilde{\rho}}
=
D_{\mathrm{KL}}(P \| \tilde{\rho})
\;\geq\; 0,
\label{eq:D_definition}
\end{equation}
provided $P \ll \tilde{\rho}$ (absolute
continuity\footnote{The symbol $\ll$ denotes absolute continuity, not
``much smaller'': $P$ assigns no probability to any region where
$\tilde\rho$ vanishes. Physically, particles are not found where the
pilot wave has zero amplitude over a finite volume. This makes the ratio
$\tilde f = P/\tilde\rho$ well defined, but it does not make $D$ finite,
which is a further condition on $P_0$ assumed at $t=0$; it is not
a bound on the size of $P$. It is a condition on the initial ensemble
$P_0$ alone, since it is preserved by the transport of $\tilde f$
(Lemma~\ref{lem:normalized_transport} below).}).
Under this hypothesis, $D = 0$ if and only if $P = \tilde{\rho}$,
equivalently $\tilde{f} = 1$. Since
$\ln \tilde{f} = \ln f + \ln N_{\mathrm{B}}$ and
$\int_{\mathcal{C}} \mathrm{d}\Omega \; P = 1$,
\begin{equation}
S_{\mathrm{e}}(t)
=
-k_{\mathrm{B}} \, D(t) + k_{\mathrm{B}} \ln N_{\mathrm{B}}(t),
\label{eq:D_S_relation}
\end{equation}
a structural contribution $-k_{\mathrm{B}} D$, measuring the
rearrangement of $P$ relative to $\tilde{\rho}$, plus an amplitude
contribution $k_{\mathrm{B}} \ln N_{\mathrm{B}}$.

When the Bohmian trajectories exist on $[0,t]$ for $P_0$-a.e.\ initial
condition (Assumption~\ref{ass:H}), the diagnostic admits the equivalent
Lagrangian representation, with $F(t;x_0)$ the trajectory-restricted
ratio of Eq.~\eqref{eq:nbody_F_definition} and
$\tilde{F} = N_{\mathrm{B}} F$,
\begin{equation}
D(t)
=
\bigl\langle \ln \tilde{F}(t;x_0) \bigr\rangle_{P_0}
=
\bigl\langle \ln F(t;x_0) \bigr\rangle_{P_0}
+ \ln N_{\mathrm{B}}(t),
\label{eq:D_lagrangian}
\end{equation}
the average being taken over the initial distribution $P_0$; its second
term is the amplitude contribution of Eq.~\eqref{eq:D_S_relation}. For the numerical examples of Sec.~\ref{sec:examples} we evaluate $D$ from the trajectory form, as described in \ref{app:numerics}, while the analytical statements rest on the spatial definition~\eqref{eq:D_definition}.

\subsection{Equilibria and Source Classification}
\label{subsec:two_channel_def}
A normalized spatial profile and an absolute Born weight answer different questions. We now make that distinction into a definition and then classify which types of source preserve each condition. The classification concerns equilibrium itself; the response of a particular transported ensemble is examined later.

\begin{definition}[Structural, amplitude, and full quantum equilibrium]
\label{def:equilibrium}
We distinguish three notions. Here $\mathcal{G}_t=\{x\in\mathcal{C}:\psi_J(x,t)\neq0\}$ denotes the nonvanishing (nonnodal) set of the guiding field, on which the ratios used below are defined (Sec.~\ref{sec:diagnostics}).
\begin{enumerate}
\item \textit{Structural equilibrium:}
$D = D_{\mathrm{KL}}(P\|\tilde{\rho}) = 0$, equivalently
$P = \tilde{\rho}$, equivalently $\tilde{f} = 1$. This is the
Valentini-type equilibrium condition, and it determines all
single-time position statistics relative to the ray of $\psi_J$.
\item \textit{Amplitude equilibrium:} the Born weight equals the
total mass of the particle ensemble, $N_{\mathrm{B}} = 1$ (since
$\int_{\mathcal{C}} \mathrm{d}\Omega\; P = 1$). This is a property of
the guiding wave relative to the particle-sector normalization.
\item \textit{Full quantum equilibrium:} $D = 0$ and
$N_{\mathrm{B}} = 1$ simultaneously, equivalently $f = 1$ on
$\mathcal{G}_t$, equivalently $P = |\psi_J|^2$ \emph{as measures}: the
two sectors agree both in shape and in total weight. This is the
recovery of \emph{Born-like statistics} within the subquantum
theory. It is \emph{not}, by itself, the recovery of standard QM
(SQM): for exogenous $J \neq 0$ the pilot-wave $\psi_J$ still
solves the ISE, $\rho = |\psi_J|^2$ still obeys the sourced
continuity equation, and the velocity field, trajectories, and
multi-time statistics all differ from their source-free
counterparts. Full SQM recovery requires in addition $J = 0$ (or
$W = 0$ on $\mathcal{G}_t$).
\textit{Sustained} full quantum equilibrium on $[0,T]$ holds if
and only if $\Gamma \equiv 0$ on $\mathcal{G}_t$ for all
$t \in [0,T]$ (given $f(0) = 1$ on $\mathcal{G}_0$); the wave sector
is then norm-conserving on $\mathcal{G}_t$.
\end{enumerate}
\end{definition}

\begin{proposition}[Algebraic equivalence of the Born rule]
\label{prop:born_two_conditions}
Given $\int_{\mathcal{C}} \mathrm{d}\Omega\; P = 1$ (under
Assumption~\ref{ass:H}), the Born rule $P = |\psi_J|^2$ is equivalent
to the conjunction of structural and amplitude equilibrium:
\begin{equation}
f = 1
\iff
D = 0 \;\text{ and }\; N_{\mathrm{B}} = 1.
\label{eq:born_rule_two_conditions}
\end{equation}
\end{proposition}
\begin{proof}
$f = 1$ means $P = \rho$, so $\int P = \int \rho$, which together with
$\int P = 1$ gives $N_{\mathrm{B}} = 1$. Then
$\tilde{\rho} = \rho = P$, so $D = 0$. Conversely, $D = 0$ gives
$P = \tilde{\rho} = \rho/N_{\mathrm{B}}$, and $N_{\mathrm{B}} = 1$
gives $\rho = \tilde{\rho}$, so $P = \rho$, i.e., $f = 1$.
Here $D=0$ means $P=\tilde\rho$ by the equality case of Gibbs'
inequality, which applies since $P \ll \tilde\rho$.
\end{proof}

With the spatial fluctuation of the source rate
\begin{equation}
\hat{\Gamma} := \Gamma - \langle\Gamma\rangle_\rho,
\label{eq:Gamma_hat}
\end{equation}
the normalized quantities obey the following transport identities.

\begin{lemma}[Normalized transport]
\label{lem:normalized_transport}
Let $D(0) < \infty$ and let $J$ be regular in time. Under
Assumptions~\ref{ass:A}--\ref{ass:E}:
\begin{align}
\partial_t\tilde\rho + \nabla\!\cdot(\tilde\rho\,v^{\psi_J})
&= \hat\Gamma\,\tilde\rho
\quad\text{pointwise on }\mathcal{G}_t, \label{eq:rho_tilde_continuity}\\
\frac{\mathrm{D}\tilde f}{\mathrm{D}t} &= -\hat\Gamma\,\tilde f
\quad\text{along Bohmian trajectories},
\label{eq:ftilde_transport}\\
\frac{\mathrm{d}D}{\mathrm{d}t} &= -\langle\hat\Gamma\rangle_P
= \langle\Gamma\rangle_\rho - \langle\Gamma\rangle_P .
\label{eq:D_rate_full}
\end{align}
\end{lemma}
\begin{proof}
Differentiating $\tilde\rho=\rho/N_{\mathrm{B}}$ with
Eq.~\eqref{eq:nbody_continuity} and
$\dot N_{\mathrm{B}}=\langle\Gamma\rangle_\rho N_{\mathrm{B}}$ gives
$\partial_t\tilde\rho = [-\nabla\!\cdot(\rho v)+\Gamma\rho]/N_{\mathrm{B}}
-\tilde\rho\langle\Gamma\rangle_\rho
= -\nabla\!\cdot(\tilde\rho v)+\hat\Gamma\tilde\rho$, which is
\eqref{eq:rho_tilde_continuity}. Inserting $P=\tilde f\tilde\rho$ in
the source-free Eq.~\eqref{eq:continuity} gives
$0=\tilde\rho\,\mathrm{D}\tilde f/\mathrm{D}t
+\tilde f[\partial_t\tilde\rho+\nabla\!\cdot(\tilde\rho v)]$, and
\eqref{eq:rho_tilde_continuity} turns this into
\eqref{eq:ftilde_transport}. Finally, $D=\langle\ln\tilde F\rangle_{P_0}$
by Eq.~\eqref{eq:D_lagrangian}, and \eqref{eq:ftilde_transport} gives
$\mathrm{d}\ln\tilde F/\mathrm{d}t=-\hat\Gamma$ along each trajectory;
integrating $\mathrm{d}\ln\tilde F/\mathrm{d}t$ in time and exchanging the time
integral with the average over $P_0$ is justified by Fubini's theorem, since
Assumption~\ref{ass:E}(i) makes $\Gamma$ and $\langle\Gamma\rangle_\rho$ integrable
with respect to $P_0\otimes\mathrm{d}t$ (with Assumption~\ref{ass:H} for the
trajectories), so that $D$ is absolutely continuous in time and its derivative
exists for almost every $t$, and
$\langle\hat\Gamma\rangle_P=\langle\Gamma\rangle_P-\langle\Gamma\rangle_\rho$.
\end{proof}
Combining~\eqref{eq:D_S_relation} with~\eqref{eq:D_rate_full} and
$\dot N_{\mathrm{B}}=\langle\Gamma\rangle_\rho N_{\mathrm{B}}$ gives the
entropy-exchange formula
\begin{equation}
\frac{\mathrm{d} S_{\mathrm{e}}}{\mathrm{d} t}
=
k_{\mathrm{B}} \langle \Gamma \rangle_P .
\label{eq:nbody_entropy_rate_unnormalized}
\end{equation}

The two quantities $\hat\Gamma$ and $\langle\Gamma\rangle_\rho$ divide
sources into four classes (Table~\ref{tab:source_classification}).
An amplitude-only source has $\hat\Gamma\equiv0$ on $\operatorname{supp}P$
and $\langle\Gamma\rangle_\rho\not\equiv0$: $D$ is conserved for any
initial value $D(0)$, and only $N_{\mathrm{B}}$ evolves. A
structural-only source has $\langle\Gamma\rangle_\rho\equiv0$ and
$\hat\Gamma\not\equiv0$: $N_{\mathrm{B}}$ is conserved while $D$ is free to
evolve. A Born-preserving source has $\langle\Gamma\rangle_\rho\equiv0$ and
$\hat\Gamma\equiv0$ on $\operatorname{supp}P$, that is, $\Gamma\equiv0$ on
$\operatorname{supp}P$ together with a vanishing $\rho$-average, as holds for
instance when $\Gamma\equiv0$ on $\mathcal{G}_t$; both quantities are then
conserved. The condition $\Gamma\equiv0$ on $\operatorname{supp}P$ alone is not
enough, because $\Gamma$ may be nonzero where $\rho>0$ and $P=0$ and then
$\langle\Gamma\rangle_\rho\neq0$. A two-channel source activates both
channels. The classes record which channel a source is able to drive; whether
$\dot D$ is nonzero at a given instant is decided by the difference
$\langle\Gamma\rangle_\rho-\langle\Gamma\rangle_P$, so that
$\hat\Gamma\not\equiv0$ does not by itself guarantee $\dot D\neq0$.
When $D<\infty$, absolute continuity $P\ll\tilde\rho$ and continuity of
$\rho$ make $P_t$ concentrated on $\mathcal{G}_t$, so all conditions are
imposed on $\mathcal{G}_t$ and, where stated up to null sets, relative to
$P_t|_{\mathcal{G}_t}$.
\begin{table}[t]
\caption{Classification of sources by channel.}
\label{tab:source_classification}
\centering
\begin{tabular}{l|cc}
\toprule
& $\hat\Gamma \equiv 0$ on $\operatorname{supp}P$ &
$\hat\Gamma \not\equiv 0$ \\
\midrule
$\langle\Gamma\rangle_\rho \equiv 0$ &
Born-preserving & Structural-only \\
$\langle\Gamma\rangle_\rho \not\equiv 0$ &
Amplitude-only & Two-channel \\
\bottomrule
\end{tabular}
\end{table}

\begin{proposition}[Conditions for maintaining equilibrium]
\label{prop:maintenance}
Under Assumptions~\ref{ass:A}, \ref{ass:H}, and \ref{ass:E}, the three
equilibria defined above are maintained exactly under
the following conditions.
\begin{enumerate}
\item \textit{Structural equilibrium ($D \equiv 0$).}
Let $D(0) = 0$ and $P_0 \ll \tilde\rho_0$.
Then $D \equiv 0$ on $[0,T]$ if and only if, for a.e.\ $t$,
$\hat{\Gamma} = 0$ $P_t$-almost everywhere,
i.e., $\Gamma$ is spatially uniform on $\operatorname{supp} P_t$;
equivalently $W = w(x,t) + \im\hbar\gamma(t)/2$ with $w$ real
and $\gamma$ spatially uniform there.
\item \textit{Amplitude equilibrium ($N_{\mathrm{B}} \equiv 1$).}
Let $N_{\mathrm{B}}(0) = 1$. Then $N_{\mathrm{B}} \equiv 1$ on
$[0,T]$ if and only if $\langle\Gamma\rangle_\rho \equiv 0$ on
$[0,T]$. The source redistributes structural nonequilibrium
without changing the global norm.
\item \textit{Full quantum equilibrium ($f \equiv 1$).}
Let $f(0) = 1$ on $\mathcal{G}_0$ and $N_{\mathrm{B}}(0) = 1$. Then
$f \equiv 1$ on $\mathcal{G}_t$ for all $t \in [0,T]$ and
$N_{\mathrm{B}} \equiv 1$ if and only if $\Gamma \equiv 0$ on
$\mathcal{G}_t$ for all $t \in [0,T]$, i.e., $J = w\psi_J$ with
$w$ real on $\mathcal{G}_t$.
\end{enumerate}
Part~(i) is stated $P_t$-almost everywhere, because $D$ compares $P$
with $\tilde\rho$ only where $P$ lives; part~(iii) is stated on all of
$\mathcal{G}_t$, because it also constrains $\rho$ away from
$\operatorname{supp}P_t$. Under $D<\infty$ the condition of part~(iii)
implies that of part~(i), but not conversely.
\end{proposition}
\begin{proof}
The three parts follow from the transport structure along trajectories, and each is argued in turn.
\begin{enumerate}
\item Initially $\tilde f \equiv 1$. By
Lemma~\ref{lem:normalized_transport}, Eq.~\eqref{eq:ftilde_transport},
$\tilde f$ stays $1$ along trajectories if and only if
$\hat\Gamma = 0$ along them, i.e., $P_t$-almost everywhere
(Assumption~\ref{ass:H}). Since $D = 0$ iff $\tilde f = 1$
$P$-a.e., this is the claim.
\item Immediate from
$\dot N_{\mathrm{B}} = \langle\Gamma\rangle_\rho N_{\mathrm{B}}$,
Eq.~\eqref{eq:nbody_NB_rate}.
\item Only if: If $f \equiv 1$ on $\mathcal{G}_t$, then
$\mathrm{D}f/\mathrm{D}t = 0 = -\Gamma f$ on $\mathcal{G}_t$, so
$\Gamma = 0$ there.
If: suppose $\Gamma = 0$ on $\mathcal{G}_t$. Then $\rho$ and $P$
solve the same source-free continuity equation with the same velocity
field. Moreover $f(0)=1$ on $\mathcal{G}_0$ gives $P_0=\rho_0$ there, and
$N_{\mathrm{B}}(0)=\int\rho_0 = 1 = \int P_0$ forces
$P_0(\mathcal{C}\setminus\mathcal{G}_0)=0$, so the two share the same
initial data. Uniqueness of transport along the trajectories
(Assumptions~\ref{ass:A} and~\ref{ass:H}) gives $\rho = P$ on
$\mathcal{G}_t$; in particular
$N_{\mathrm{B}} = \int \rho = \int P = 1$. Finally
$\operatorname{Im}(J/\psi_J) = 0$ on $\mathcal{G}_t$ means $J = w\psi_J$
with $w$ real.
\end{enumerate}
\end{proof}

\begin{remark}[Which source components control which channels]
\label{rem:source_channels}
With the complex source ratio $W = J/\psi_J$ of
Eq.~\eqref{eq:W_definition}, three projections can be identified:
\begin{itemize}
\item The spatial gradient $\nabla\operatorname{Re} W$ enters the
modified Hamilton--Jacobi equation~\eqref{eq:nbody_hj} and changes
the phase $S_J$, hence the guidance velocity and the trajectories.
A spatially uniform $\operatorname{Re} W$ only adds a global
phase. Additionally, nonuniform $\Gamma$ generically deforms
$R_J$ through the ISE, which changes $Q^{\mathrm{B}}_J$ and thus
the trajectories indirectly; the trajectory channel is therefore
controlled by $\nabla\operatorname{Re} W$ \emph{plus} the induced
change in $Q^{\mathrm{B}}_J$.
\item The global average $\langle\Gamma\rangle_\rho =
(2/\hbar)\langle\operatorname{Im} W\rangle_\rho$ controls the
Born norm $N_{\mathrm{B}}(t)$.
\item The spatial fluctuations
$\hat{\Gamma} = \Gamma - \langle\Gamma\rangle_\rho$
control the structural diagnostic $D$.
\end{itemize}
The three projections are independent as degrees of freedom of the
pair $(\operatorname{Re} W, \operatorname{Im} W)$, but the resulting
dynamics are coupled through the shared field $\psi_J$, since the
velocity, the rate $\Gamma$, and the amplitude $R_J$ all depend on the
same solution. In particular, the structural channel feeds into the
trajectory channel through the deformation of $R_J$ and hence of
$Q^{\mathrm{B}}_J$.

\end{remark}

\begin{remark}[State-proportional sources and the amplitude channel]
\label{rem:proportional_channels}
For the state-proportional source $J = \kappa(t)\psi_J$ with
$\kappa = \kappa_{\mathrm{r}} + \im\kappa_{\mathrm{i}}$, the ratio
$W = \kappa(t)$ is spatially uniform, so $\nabla\operatorname{Re} W = 0$
and $\hat\Gamma = 0$. Only the amplitude channel is active, and the
trajectories coincide with the source-free ones.
\end{remark}

\section{Solution of the Main Equations of Subquantum System Theory}
\label{sec:formalism}
By the one-way hierarchy~\eqref{eq:causal_hierarchy}, the solution proceeds sequentially, first the field sector and then the particle sector; the function spaces and notation used are collected in \ref{app:math}.

\subsection{The Field Sector: Duhamel Solution of the ISE}
\label{sec:duhamel}
The two-channel rate laws hold for a general guiding field, but their use in a concrete system begins with a field solution. Because the ISE~\eqref{eq:ise} is linear and inhomogeneous on $\mathcal H=L^2(\mathcal C)$, its solution has the Duhamel representation \cite{cazenave2003semilinear}

\begin{equation}
\psi_J(x,t)
=
U(t,0)\,\psi_0(x)
-
\frac{\im}{\hbar}
\int_0^t \mathrm{d}s\;
U(t,s)\,J(x,s),
\label{eq:duhamel}
\end{equation}
where $U(t,s) := e^{-\im H(t-s)/\hbar}$ is the unitary propagator generated by the time-independent
self-adjoint Hamiltonian $H$, whose existence is guaranteed by
Stone's theorem \cite{reed1980functional_1}. The Duhamel map is
\emph{jointly linear} in $(\psi_0, J)$: the joint rescaling
$(\psi_0, J) \to (c\psi_0, cJ)$ maps solutions to solutions, which is
the origin of the joint-rescaling invariance stated in
Remark~\ref{rem:joint_rescaling}.

The solution~\eqref{eq:duhamel} decomposes naturally into two
contributions, $\psi_J=\psi_{\mathrm{hom}}+\psi_{\mathrm{inh}}$, where the \emph{homogeneous part}
$\psi_{\mathrm{hom}}(x,t) := U(t,0)\psi_0(x)$ is the free
Schr\"odinger evolution of the initial data, and the
\emph{inhomogeneous part}
\begin{equation}
\psi_{\mathrm{inh}}(x,t)
:=
-\frac{\im}{\hbar}
\int_0^t \mathrm{d}s\;
U(t,s)\,J(x,s)
\label{eq:psi_inh}
\end{equation}
is the cumulative response of the system to the source history
$\{J(\cdot,s)\}_{s\in[0,t]}$. The unitary propagator $U(t,s)$ acts on
both components: it evolves the initial datum forward in the
homogeneous part, and it propagates each instantaneous source
contribution $J(\cdot,s)$ from the emission time $s$ to the
observation time $t$ in the inhomogeneous part. The standard sufficient conditions for the Duhamel integral to be well defined, and the regularity that the guidance equation requires (Assumption~\ref{ass:A}), are collected in \ref{rem:duhamel_wellposed}.

\subsection{Exogenous vs.\ Endogenous Sources and the Physical Status
of $N_{\mathrm{B}}$}
\label{subsec:source_types}
The Duhamel formula exposes a distinction that is invisible if one looks only at the value of $J/\psi_J$ at a single time. Is the source prescribed independently of the guiding state, or is it generated by the state itself? The answer determines which rescalings describe the same dynamics, and hence what the amplitude of the guiding wave can mean. We distinguish the two specifications before examining scale covariance.

\begin{definition}[Source classification by dependence on state]
\label{def:source_types}
Fix $H$. A source specification is:
\begin{enumerate}
\item \textit{Exogenous:} a single prescribed field $J(x,t)$
chosen independently of $\psi_0$.
\item \textit{Endogenous:} a source $J = \mathcal{K}[\psi_J]$ with
$\mathcal{K}$ a map fixed in advance for all initial data, and
$\mathcal{K}[0] = 0$.
\item \textit{Mixed affine:} a source $J = J_0 +
\mathcal{K}[\psi_J]$ with $J_0 \not\equiv 0$ prescribed and
$\mathcal{K}$ as in~(ii).
\end{enumerate}
Note that any specific solution $\psi_J$ allows writing
$J = (J/\psi_J)\psi_J$ post-hoc; the endogenous definition requires
$\mathcal{K}$ to be fixed in advance for all $\psi_0$. A source with
$\mathcal{K}[0] \neq 0$ is mixed affine, not endogenous.
\end{definition}

The operational test is whether $\psi\equiv0$ solves the ISE: it does for
an endogenous source with $\mathcal{K}[0]=0$ and does not for an
exogenous or mixed affine source with $J_0\neq0$. A source
$J=W(x,t)\psi_J$ with prescribed $W$ is endogenous (linear in $\psi_J$).

In source-free quantum mechanics the evolution is linear, so the
rescaling $\psi_0\mapsto c\psi_0$ changes only the normalization of the
same ray and leaves the guidance velocity, $\tilde\rho$ and $D$ untouched.
An independent exogenous source supplies a second amplitude and with it a
reference against which the amplitude of the guiding wave can be measured:
at fixed $J$ the solution $\psi_J=cU\psi_0+\psi_{\mathrm{inh}}$ no longer
scales as a whole, so the rescaling ceases to be a symmetry and can alter
the velocity, $D$ and the trajectories. This is a statement about the
dynamics, and it rests only on the linearity of the Duhamel map. Treating
the absolute amplitudes of $\psi_J$ and $J$ as physical is an ontological
premise of the present theory, in which both are real fields, and not a
consequence of Proposition~\ref{prop:scale_covariance}; the proposition
identifies instead a dynamical role for the amplitude of the guiding wave
relative to an independently specified source. Rescaling the initial guiding
field at fixed source thus tests whether that relative amplitude has
dynamical consequences, and Proposition~\ref{prop:scale_covariance}
separates sources accordingly: for linear endogenous sources rescaling $\psi_J$ is a
symmetry of the guidance dynamics and of the normalized diagnostics, and
for exogenous and mixed affine sources it is not. ``Scale covariance'' here
means invariance of $v^{\psi_J}$, $\tilde\rho$, $D$, $W$ and $\Gamma$
under $\psi_J\mapsto c\psi_J$, and the proof is given in
\ref{app:scale_proof}.

\begin{proposition}[Scale covariance and the physical status of
$N_{\mathrm{B}}$]
\label{prop:scale_covariance}
Assume that the ISE with the source in question is well posed, so that its solutions are unique; this is a hypothesis on the effective generator for endogenous and mixed sources, which is not the free propagator $U$. The scale properties are as follows.
\begin{enumerate}
\item \textit{Endogenous, linear generator.}
If $\mathcal{K}$ is $\mathbb{C}$-linear, then
$\psi_J \mapsto c\psi_J$ ($c \in \mathbb{C}\setminus\{0\}$) maps
solutions to solutions. Then $v^{\psi_J}$, $\tilde{\rho}$, $D$,
$W$, $\Gamma$, and $\langle\Gamma\rangle_\rho$ are all invariant,
while $N_{\mathrm{B}}(t) \mapsto |c|^2 N_{\mathrm{B}}(t)$. Only the
relative evolution $N_{\mathrm{B}}(t)/N_{\mathrm{B}}(0)$ enters the
dynamics, and the absolute scale of $N_{\mathrm{B}}$ plays no dynamical role. If
$\mathcal{K}$ is only positively homogeneous, the same holds for
$c > 0$.
\item \textit{Exogenous and mixed affine.}
Let $\mathcal{K}$ be $\mathbb{C}$-linear (with $\mathcal{K}=0$ in the
exogenous case), $J_0 \not\equiv 0$, and let $\psi_{\mathrm{inh}}$ be
the solution with zero initial data. If $\psi_J$ is the solution for
data $\psi_0$ and $\psi_J^{(c)}$ for data $c\psi_0$, then
$\psi_J^{(c)} = c\,\psi_J + (1-c)\,\psi_{\mathrm{inh}}
= [c + (1-c)r]\,\psi_J$ with $r := \psi_{\mathrm{inh}}/\psi_J$.
In particular, for $c\neq1$, $\psi_J^{(c)} \neq c\psi_J$ unless
$\psi_{\mathrm{inh}} = 0$.
\item \textit{Velocity criterion.}
Rescaling the data $\psi_0 \to c\psi_0$ at fixed $J$ ($c\neq1$) leaves
the guidance velocity unchanged on the common nonnodal domain $\{\psi_J\neq0,\ \psi_J^{(c)}\neq0\}$, where both velocities are defined, if and only if
\begin{equation}
\operatorname{Im}\!\left[
\frac{(1-c)\nabla r}{c + (1-c)r}
\right] = 0.
\label{eq:velocity_invariance_condition}
\end{equation}
This holds for all $c > 0$ if and only if $r$ is real-valued or
spatially constant, and for all complex $c$ with
$\operatorname{Im} c \neq 0$ only if $r$ is spatially constant. When the
condition fails, the trajectories depend on the amplitude of $\psi_0$
relative to $J$, which is therefore dynamical data.
\end{enumerate}
\end{proposition}

\begin{remark}[Joint rescaling and what $N_{\mathrm{B}} = 1$ means]
\label{rem:joint_rescaling}
By the joint linearity of the Duhamel map in $(\psi_0, J)$, the joint
rescaling $(\psi_0, J) \to (c\psi_0, cJ)$ maps solutions to solutions and
preserves the normalized particle-sector dynamics: the structural
diagnostic $D$, the velocity field $v^{\psi_J}$, the normalized ratio
$\tilde{f}$, the local rate $\Gamma$ and the growth rate
$\mathrm{d}\ln N_{\mathrm{B}}/\mathrm{d}t = \langle\Gamma\rangle_\rho$ are
invariant, while $N_{\mathrm{B}} \to |c|^2 N_{\mathrm{B}}$ and
$f \to f/|c|^2$. The strength of the source relative to the initial field is
unchanged by this rescaling, so the condition $N_{\mathrm{B}} = 1$ cannot
fix it: it fixes the overall scale of the pair $(\psi_0, J)$, that is, the
calibration of the field amplitude against the particle ensemble, whose total
weight is one, while the relative source strength is a separate datum. Two readings are consistent with this
mathematics. If the amplitudes of $\psi_J$ and $J$ are conventional, jointly
rescaled configurations describe one physical state and $N_{\mathrm{B}} = 1$
is a choice of units; the amplitude channel then reduces to the invariant
growth rate $\langle\Gamma\rangle_\rho$ together with the relative source
strength. If, as in the ontology adopted here, $\psi_J$ and $J$ are real
fields with absolute amplitudes, jointly rescaled configurations are distinct
states with identical normalized particle dynamics, and $N_{\mathrm{B}}$
measures the weight of the guiding field against the ensemble. The algebraic
statements are valid independently of whether the common field--source
amplitude is regarded as conventional or physical, and the latter reading is
adopted here. Measurements restricted to the normalized single-time particle
distribution $P(\cdot,t)$ cannot determine the absolute Born weight, and
joint rescaling leaves the normalized particle-sector dynamics unchanged, but
this restricted observational equivalence is not a no-go theorem: a
calibrated exogenous source together with information on the normalized
guiding field allows $N_{\mathrm{B}}$ to be reconstructed in principle,
under the idealized premise of real and comparable fields, as sketched in
\ref{app:measure}. For mixed affine sources the statement presupposes the
fixed-generator hypotheses of Proposition~\ref{prop:scale_covariance}.
\end{remark}

Equation~\eqref{eq:velocity_invariance_condition} has a simple phase interpretation. At fixed source the rescaled field is $\psi_J^{(c)}=[c+(1-c)r]\psi_J$, and the guidance velocity depends on the spatial phase gradient of the multiplier $c+(1-c)r$, not on its modulus, so that rescaling the initial field changes the velocity only when the multiplier acquires a spatially varying phase. For positive rescalings a real $r$ changes only the local magnitude away from nodes, and a spatially constant $r$ produces only a global factor; in either case the velocity is unchanged. This velocity-level invariance is weaker than invariance of the Born density and of the two diagnostics, which must not be inferred from it, and a rescaling may also create new nodes, so the comparison is confined to the region where both velocity fields are defined.

\begin{remark}[Null start]
\label{rem:null_start}
For exogenous $J\not\equiv0$ on $[0,t]$, the initial condition
$\psi_0=0$ generates $\psi_J(t)=\psi_{\mathrm{inh}}(t)$, which is nonzero
for small $t>0$ when $J(0)$ is continuous and nonzero and generically
thereafter, although Duhamel contributions can cancel at isolated times: for
$H\varphi=E\varphi$ and $J=j\,e^{-\im Es/\hbar}\cos(2\pi s/T)\varphi$ one finds
$\psi_J(t)=-(\im jT/2\pi\hbar)\,e^{-\im Et/\hbar}\sin(2\pi t/T)\varphi$, which
vanishes at $t=T$. No
endogenous theory with Lipschitz $\mathcal{K}$ and $\mathcal{K}[0]=0$
can do this, since $\psi\equiv0$ is then a solution and uniqueness
forbids leaving it; a null start is therefore a distinctive signature of
exogenous sources. For the structural diagnostic
$D(0^+)=D_{\mathrm{KL}}(P_0\,\|\,|J(t_0)|^2/\|J(t_0)\|_{L^2}^2)$ to be
defined one needs $J(t_0)\in L^2$ and nonzero, $P_0$ absolutely
continuous with respect to $|J(t_0)|^2/\|J(t_0)\|_{L^2}^2$, and
$J\neq0$ on $\operatorname{supp}P_0$, so that the initial velocity
$(\hbar/m)\operatorname{Im}(\nabla J/J)$ is finite. At $t=0$ itself the
guidance velocity is undefined because $\psi_0=0$, so the particle sector is
started at a positive time $t_0^+$ with the initial ensemble prescribed there.
\end{remark}

\begin{remark}[Exact rescaling for state-proportional sources]
\label{rem:exact_rescaling_2}
A particularly transparent endogenous case is the
state-proportional source $J = \kappa(t)\psi_J$ with
$\kappa(t) = \kappa_{\mathrm{r}}(t) + \im\kappa_{\mathrm{i}}(t)$ a
complex scalar. Then $W = \kappa(t)$, and the ISE becomes $\im\hbar\partial_t\psi_J=(H+\kappa_{\mathrm{r}}+\im\kappa_{\mathrm{i}})\psi_J$. If $\psi_0^{\mathrm{free}}(t)$ solves the source-free equation with
the same initial data, the solution is
\begin{equation}
\psi_J(t)
=
\exp\!\left[\frac{1}{\hbar}\int_0^t \kappa_{\mathrm{i}}(s)\,\D s\right]
\exp\!\left[-\frac{\im}{\hbar}\int_0^t \kappa_{\mathrm{r}}(s)\,\D s
\right]\psi_0^{\mathrm{free}}(t).
\label{eq:exact_rescaling}
\end{equation}
The first factor is a real, time-dependent rescaling; the second is a
global phase. Both prefactors are functions of $t$ alone. Writing
$\psi_0^{\mathrm{free}} = R^{\mathrm{free}}e^{\im S^{\mathrm{free}}/\hbar}$
and denoting by $A(t)$ the real prefactor, the polar decomposition of
$\psi_J$ reads $R_J = A(t)R^{\mathrm{free}}$ and
$S_J = S^{\mathrm{free}} + \hbar\varphi(t)$ with
$\varphi(t) := -\hbar^{-1}\int_0^t \kappa_{\mathrm r}(s)\,\D s$. Since
$\varphi$ is spatially uniform, $\nabla_i[\hbar\varphi(t)] = 0$, and
therefore $\nabla_i S_J = \nabla_i S^{\mathrm{free}}$ for every $i$.
Consequently the velocity field
$\mathbf{v}^{\psi_J} = \nabla S_J/m$ and the Bohmian trajectories
coincide with the source-free case. The Born norm evolves as $N_{\mathrm{B}}(t)=N_{\mathrm{B}}(0)\exp[(2/\hbar)\int_0^t\kappa_{\mathrm{i}}(s)\,\D s]$, and the local source rate is $\Gamma = 2\kappa_{\mathrm{i}}/\hbar$,
spatially uniform. Thus the proportional source drives amplitude
dynamics without affecting the structural or trajectory channels.
\end{remark}

\begin{remark}[Physical significance of the amplitude diagnostic]
\label{rem:NB_physical_significance}
The structural equilibrium condition $D = 0$ yields
$P = \tilde{\rho} = |\psi_J|^2/N_{\mathrm{B}}$, a
\emph{Born-like} distribution for single-time position statistics.
However, this is a Born-like distribution \emph{within the
subquantum theory}, not a recovery of standard QM (SQM). Three
points qualify the identification.
\begin{enumerate}
\item \textit{The wave sector is not SQM.} The pilot-wave
$\psi_J$ satisfies the ISE
$\im\hbar\partial_t\psi_J = (H + W)\psi_J$, not the
source-free Schr\"odinger equation. In SQM, $\rho = |\psi|^2$ is
conserved; here, $\rho = |\psi_J|^2$ satisfies the modified
continuity equation with source $\Gamma\rho$, and only
$\tilde{\rho} = \rho/N_{\mathrm{B}}$ is normalized. The velocity
field $v^{\psi_J}$, the trajectories, and the multi-time statistics
all differ from their source-free counterparts whenever $\Gamma$ is not
a spatial constant. Exact reduction to the source-free field equation
requires $J = 0$ (or $W = 0$ on $\mathcal{G}_t$). A real and spatially
uniform $W = w(t)$ adds only a global phase and is ray-equivalent to SQM
at the level of normalized predictions, whereas the two-channel criterion
$D = 0$, $N_{\mathrm{B}} = 1$ is stricter than both, since it also fixes
the Born weight.
\item \textit{The amplitude relative to the source is dynamical.} In SQM
only the ray $[\psi]$ matters. With an exogenous $J$ the source breaks
$U(1)$ phase symmetry and supplies both a phase and an amplitude
reference. The local rate $\Gamma = (2/\hbar)\operatorname{Im}(J/\psi_J)$
depends on the amplitude of $\psi_J$ relative to $J$, and the phase of
$\psi_J$ is fixed relative to that of $J$, so the ratio of the field
amplitude to the source strength is dynamical data
(Proposition~\ref{prop:scale_covariance}). This ratio has dimension of
inverse energy, since $J$ carries units of energy times wavefunction; a
dimensionless measure is
$\chi_T = (\hbar\|\psi_0\|_2)^{-1}\int_0^T\|J(s)\|_2\,\mathrm{d}s$.
\item \textit{Amplitude and structure are coupled in the exogenous case.}
For fixed exogenous $J$, a change in the initial amplitude $|\psi_0|$
alters $\Gamma$, which drives $D$ and modifies the trajectories: the
amplitude of the guiding wave acts on the structural channel through the
ratio $|J|/|\psi_0|$. See Example~HO.4 (Sec.~\ref{sec:feedback}) for the
explicit realization.
\end{enumerate}
\end{remark}

\subsection{The Particle Sector: Transport of the Probability
Distribution}
\label{sec:particle_sector}
Once the guiding field has been obtained, the remaining problem is to follow the particle ensemble through its velocity field. The distinction between an exact field solution and a generally nontrivial particle flow is important: a simple $\psi_J$ need not produce simple trajectories. Starting at $x_0\in\operatorname{supp}P_0$, a characteristic $q(t;x_0)$ obeys

\begin{equation}
\frac{\mathrm{d}q}{\mathrm{d}t}(t; x_0)
=
v^{\psi_J}(q(t; x_0),\,t),
\qquad
q(0; x_0) = x_0 .
\label{eq:guidance_ODE}
\end{equation}
Assumption~\ref{ass:A} guarantees a unique local solution on its maximal interval $I(x_0)$, not necessarily across the full observation interval: a path may encounter a node or escape to infinity \cite{berndl1995global_existence}. The continuum conclusions rely on Assumption~\ref{ass:H} where complete trajectories are required. The nonequilibrium ratio $f=P/\rho$, and its normalized counterpart $\tilde f=N_{\mathrm B}f$, are transported along the characteristics. Writing $F(t;x_0)=f(q(t;x_0),t)$ as in Eq.~\eqref{eq:nbody_F_definition}, their evolution integrates to

\begin{equation}
F(t;x_0)
=
f_0(x_0)\,
\exp\!\left[
-\int_0^t \mathrm{d}s\;
\Gamma(q(s;x_0),\,s)
\right],
\label{eq:F_solution}
\end{equation}
Here $f_0(x_0)=P_0(x_0)/\rho(x_0,0)$, for initial positions with $\psi_0(x_0)\ne0$. For a null start the construction is restarted at $t_0^+$ using $\tilde f$ (Remark~\ref{rem:null_start}). The integral samples $\Gamma$ along the moving path; except in special cases such as the uniform rate of HO.2 (Sec.~\ref{sec:ho2}), the trajectory and $\ln F$ must therefore be evolved together, with $\mathrm d\ln F/\mathrm dt=-\Gamma(q,t)$ (Appendix~\ref{app:numerics}). The ring (Sec.~\ref{sec:ring}) admits a separate phase quadrature, and after the impulse in F$\infty$.1 (Sec.~\ref{sec:free_infinite}) the rate vanishes. HO.1, HO.3, HO.4 and the two-body system require coupled integration (Secs.~\ref{sec:ho} and~\ref{sec:twobody}).

This is where the source-extended dynamics becomes more than an affine field equation. Although $\psi_J$ depends linearly on $\psi_0$ and $J$, the guidance velocity depends on their ratio; a small source correction can change the particle paths and accumulate over time. Where the field vanishes, both velocity and source rate may lose regularity. The general characteristic identities remain conditional on Assumptions~\ref{ass:A}--\ref{ass:E}, and the effect of near-nodal passages on numerical ensemble quadrature is analyzed in \ref{app:numerics}.

\section{Solvable Driven Subquantum Systems}
\label{sec:examples}
The classification of equilibrium tells us which channels a source can excite, but not how a particle ensemble actually responds. That question requires following the guiding field and the trajectories together. We therefore proceed through a sequence of driven systems, beginning with the limits of a global description in infinite space, moving to the ring where the amplitude channel is silent, and then to oscillators and a two-body model in which the structural and amplitude responses can be compared. Table~\ref{tab:summary} summarizes the results. All sources are exogenous and prescribed, in the sense of Definition~\ref{def:source_types}.

An exact field solution does not generally give an exact ensemble solution. The guiding field is analytical, modal or semi-analytical in these models (the finite Gaussian pulse requires time quadrature), whereas the trajectories, the ratio field and the structural diagnostic usually require numerical integration. We advance the characteristic system of Eq.~\eqref{eq:main_ode} in Appendix~\ref{app:numerics}, reconstruct $P$ independently from the flow Jacobian, and evaluate $D$ over the initial ensemble. That appendix also presents the implementation, convergence and consistency tests. The point impulse and HO.2 are fully analytic, as are the ring invariant, its period and rate, and the joint-rescaling identity of HO.4; the plotted ensemble curves of F$\infty$.2, the ring, HO.1, HO.3, HO.4 and the two-body system are numerical. These trajectory and ensemble calculations cover the stated finite intervals and do not constitute a general proof of completeness.

We keep $\hbar$, the masses and frequencies explicit in the analytical formulas, but use scaled units for numerical values and figures. For HO.1--HO.4 and the two-body model we take $\ell_0=\sqrt{\hbar/m\omega}$, $t_0=1/\omega$, $E_0=\hbar\omega$, with $\hbar=m=\omega=1$ in the computations; axes therefore display $x/\ell_0$, $\omega t$, $\langle\Gamma\rangle/\omega$ and $j_0/E_0$, where $j_0$ is the energy coefficient of a normalized oscillator mode. For the ring, $\ell_*=L/2\pi$ (here $1$), $t_*=m\ell_*^2/\hbar$ and $E_*=\hbar/t_*$; F$\infty$.2 uses the common Gaussian width as $\ell_*$ with the same time and energy scales. Its $j_0$ multiplies a dimensionless spatial profile and a unit-integral time profile, and is therefore quoted in units of $\hbar\ell_*^{-1/2}$. Velocities and rates in oscillator figures are expressed in $\ell_0\omega$ and $\omega$, respectively; on the ring they are expressed in $\ell_*/t_*$ and $1/t_*$. The diagnostics $D$, $N_{\mathrm B}$, $\ln F$, $t/T$, $\omega t$, $t/t_*$ and the plotted errors in $D$ or logarithmic ratios are dimensionless. Further numerical conventions are collected in \ref{app:numerics}.

\subsection{Infinite-Space Free Particle: Green-Function Examples}
\label{sec:free_infinite}
Infinite space provides the simplest source solutions but also shows why local guidance and global equilibrium are different questions. A point impulse, and a nonzero plane-wave response, produce guiding fields outside $L^2$. Their Born weights are infinite, except at isolated off-resonant plane-wave recurrence times when $\psi_J\equiv0$ and $N_{\mathrm B}=0$. The normalized diagnostics $D$, $\tilde f$ and $S_{\mathrm e}$ are thus undefined, not zero (Remark~\ref{rem:NB_square_integrability}). The local velocity, source rate and trajectory equations nevertheless retain meaning wherever the field is nonzero.

\exhead{F$\infty$.1}{point impulse}Consider the sharpest possible drive: $J=j_0\,\delta^{(3)}(\mathbf{x})\,\delta(t-t_0)$ acting on $\psi_0=0$. For $t>t_0$, write $\tau=t-t_0$ and $r=|\mathbf{x}|$. The field is
\begin{equation}
\psi_J(\mathbf{x},t)
=
-\frac{\im j_0}{\hbar}
\left(\frac{m}{2\pi\im\hbar\tau}\right)^{3/2}
\exp\!\left[\frac{\im m r^2}{2\hbar\tau}\right],
\label{eq:point_source_psi}
\end{equation}
Its phase $S=mr^2/(2\tau)$ gives $\mathbf v=\mathbf x/\tau$: the trajectories are radial rays $\mathbf q(t)=\mathbf w(t-t_0)$ with constant $\mathbf w$. Once the impulse has passed, $\Gamma=0$ and $F$ is constant along each ray. Thus the impulse changes the local guidance, but does not subsequently drive either channel. As $\tau\to0^+$ the phase diverges away from the origin; the natural data for the impulse limit are the ray velocities $\mathbf w$, not initial positions. The field modulus is uniform in space, $|\psi_J|=(m/2\pi\hbar\tau)^{3/2}|j_0|/\hbar$. The consequent divergence of its norm, and the corresponding plane-wave exception, can be stated precisely.

\begin{proposition}[Born weight divergence for idealized sources on
$\mathbb{R}^3$]
\label{prop:NB_infinity}
For a free particle on $\mathbb{R}^3$ with $\psi_0=0$, if $J$ is a
spacetime impulse $j_0\,\delta^{(3)}(\mathbf{x}-\mathbf{x}_0)\,
\delta(t-t_0)$ or a plane wave $j_0\,e^{\im(\mathbf{k}\cdot\mathbf{x}-
\Omega t)}$ with $j_0\neq0$, then $N_{\mathrm{B}}(t)=\infty$ for every $t>t_0$ in the impulse case and for every $t>0$ at which the plane-wave amplitude $c(t)$ of the proof is nonzero, so that the normalized quantities $\tilde\rho$ and $\tilde f$ are undefined.
\end{proposition}
\begin{proof}
For the impulse and $t>t_0$, the Duhamel formula~\eqref{eq:duhamel} gives
$\psi_J=-(\im j_0/\hbar)G_0(\mathbf{x},t;\mathbf{x}_0,t_0)$, where
\[
G_0(\mathbf{x},t;\mathbf{x}',s)=\Bigl(\frac{m}{2\pi\im\hbar(t-s)}\Bigr)^{3/2}
\exp\Bigl[\frac{\im m|\mathbf{x}-\mathbf{x}'|^2}{2\hbar(t-s)}\Bigr]
\]
is the kernel of $U(t,s)$, whose modulus does not depend on $\mathbf{x}$; hence
$N_{\mathrm{B}}=(|j_0|/\hbar)^2(m/2\pi\hbar\tau)^3\int\mathrm{d}^3x=\infty$.
For the plane wave, $e^{\im\mathbf{k}\cdot\mathbf{x}}$ is an eigenfunction
of $H_0$, so $\psi_J=c(t)e^{\im\mathbf{k}\cdot\mathbf{x}}$ with, for $\Omega\neq\omega_k=\hbar|\mathbf{k}|^2/2m$, $c(t)=j_0\,[e^{-\im\Omega t}-e^{-\im\omega_kt}]/[\hbar(\Omega-\omega_k)]$ and, on resonance, $c(t)=-(\im j_0t/\hbar)e^{-\im\Omega t}$; hence $N_{\mathrm{B}}=|c(t)|^2\int\mathrm{d}^3x=\infty$ whenever $c(t)\neq0$. Off resonance $c$ vanishes at the recurrence times $t=2\pi n/|\Omega-\omega_k|$, $n=1,2,\dots$, where the field vanishes everywhere and $N_{\mathrm{B}}=0$; the normalized quantities are then undefined because the norm is zero, not infinite.
\end{proof}
For a nonzero impulse response the divergence comes from instantaneous spreading with uniform modulus; a plane wave is delocalized from the outset. These solutions belong to $\mathcal{S}'$ rather than $L^2$ (\ref{rem:duhamel_wellposed}). At a plane-wave recurrence time the field instead vanishes and the guidance velocity is undefined. To obtain finite global diagnostics we must regularize the source.

\exhead{F$\infty$.2}{regularized impulse and the jump rule}A finite spatial profile removes the point-source singularity, but a sudden kick presents a new question: what happens to the particle-to-field ratio while the field changes instantaneously? Take $J=j_0\,g(\mathbf{x})\,\delta(t-t_0)$ with a Gaussian $g$. For $t>t_0$, $\psi_J=U(t,0)\psi_0-(\im j_0/\hbar)U(t,t_0)g$, so that the field jumps by $\psi_+=\psi_--(\im/\hbar)j_0g$. The particle positions cannot jump with it. Hence $P$ is continuous across the kick while $\rho$ is not, and their ratio obeys the exact rule

\begin{equation}
F_+ = F_-\,\frac{|\psi_-|^2}{|\psi_+|^2},
\label{eq:jump_rule}
\end{equation}
Equivalently, $F_+=F_-/|1-\im z|^2$, where $z=j_0g/(\hbar\psi_-)$ and $|1-\im z|^2=1+|z|^2+2\operatorname{Im}z$. After the kick $\Gamma=0$, so the new value of $F$ is transported unchanged along the ensuing source-free trajectories. The jump is therefore a change in the relation between $P$ and $\rho$, not an instantaneous displacement of particles.

\begin{figure}[t]
\centering
\includegraphics[width=\textwidth]{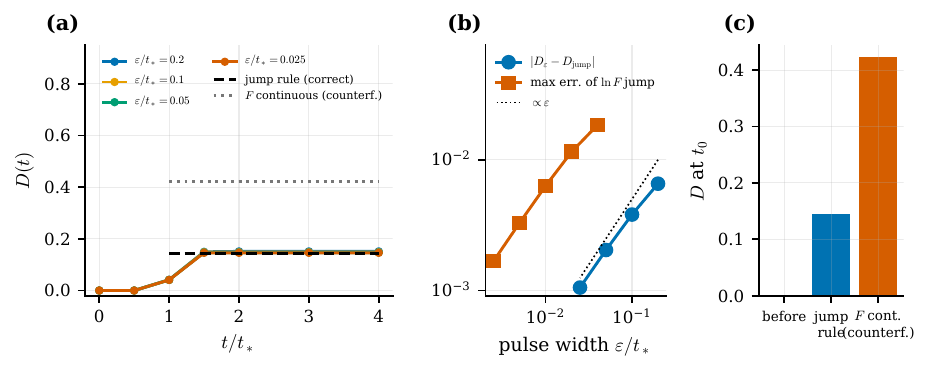}
\caption{Example F$\infty$.2 (one-dimensional free particle; times in $t_*$, scaled units of Sec.~\ref{sec:examples}). (a)~$D(t)$ from the direct integration
through pulses of width $\varepsilon$, from the jump-rule prediction and from
the counterfactual with $F$ held continuous across the pulse (gray dotted line), which is not a valid solution. (b)~Convergence to the prediction in $\varepsilon$:
difference in the final $D$ and maximum pointwise error of the jump in $\ln F$.
(c)~$D$ immediately before and after the instantaneous pulse.}
\label{fig:jump}
\end{figure}

We test the rule by resolving a pulse of small but finite duration. In one dimension we take $\psi_0=\pi^{-1/4}e^{-x^2/2+\im x}$, $P_0=|\psi_0|^2$, $g=e^{-x^2/2}$, $j_0=0.4\,\hbar\,\ell_*^{-1/2}$, and a unit-integral Gaussian pulse $h_\varepsilon$ of width $\varepsilon$ centered at $t_0=1\,t_*$. Direct characteristic integration through the pulse approaches the exact jump prediction linearly in $\varepsilon$ (Fig.~\ref{fig:jump}); the independent flow-Jacobian route to $D$ agrees with integration of $\Gamma$ (\ref{app:numerics}). With the source switched off, $D$ remains at $\nv{jumpDjump}$. Incorrectly holding $F$ continuous would instead give $\nv{jumpDno}$, about three times as large. The difference is not a numerical detail: the discontinuity of $F$ is the structural response to an instantaneous source.

\label{sec:num_jump}

\exhead{F$\infty$.3}{finite-duration pulse}A pulse $J=j_0\,g(\mathbf{x})h(t)e^{-\im\Omega t}$ links the sudden kick to a continuously driven field. As $h\to\delta$ it approaches the impulse, while $h\to\mathrm{const}$ gives the monochromatic limit. During the drive $\Gamma$ need not vanish and both the trajectory and amplitude responses may be active; after switch-off the system evolves source-free from its altered state.

This family marks the boundary of the present calculation. A finite Born weight requires square-integrable $g$ and $\psi_0$, whereas the idealized point and plane-wave solutions have no normalized global diagnostics. We do not compute a global $D$ for F$\infty$.3. The ring provides a controlled finite-volume setting in which the effect of a sustained source can be followed analytically and numerically.

\subsection{Finite-Volume Free Particle: The Ring}
\label{sec:ring}
\exhead{Ring}{trapped and untrapped regimes}Can the shape of the particle ensemble change while the norm of the guiding field remains fixed? The ring isolates this possibility. Compactifying the free-particle domain makes the guiding state square-integrable and its spatial dynamics periodic in a moving frame. On a ring of length $L$, the boundary conditions $\psi(x+L,t)=\psi(x,t)$ give a self-adjoint free Hamiltonian with modes $\varphi_k=L^{-1/2}e^{\im kx}$, $k=2\pi n/L$, and energies $\hbar\omega_k=\hbar^2k^2/2m$. A steady monochromatic drive $J=j\,e^{\im(qx-\Omega t)}$, with $q$ a ring mode and switched on at $t=-\infty$, yields

\begin{equation}
\psi_J(x,t) = c_0\,\varphi_k\,e^{-\im\omega_k t}
+ c_q\,\varphi_q\,e^{-\im\Omega t}.
\label{eq:ring_psi}
\end{equation}
The modes are orthonormal, so $N_{\mathrm B}=|c_0|^2+|c_q|^2$ is constant and $\langle\Gamma\rangle_\rho=0$. The amplitude channel is silent even though the source remains present. We choose $P_0=1/L$, a uniform initial ensemble, while $\rho_0$ has interference fringes; hence $D(0)>0$ and $N_{\mathrm B}=1$. The question is how this pre-existing structural departure evolves.

The relative phase $\theta=(q-k)(x-ut)$, where $u=(\Omega-\omega_k)/(q-k)$, moves with speed $u$. Since both $v=(\hbar/m)\operatorname{Im}(\partial_x\psi_J/\psi_J)$ and $\rho=|\psi_J|^2$ depend on this phase alone, the motion has a conserved quantity in the moving frame.

\begin{proposition}[Ring invariant]
\label{prop:ring_invariant}
For the steady driven state~\eqref{eq:ring_psi}, the combination
$F\rho(v-u)$ is constant along every Bohmian trajectory, so that
\begin{equation}
F(t)
=
\frac{P_0(v_0 - u)}{\rho(\theta(t))\,(v(\theta(t)) - u)}.
\label{eq:ring_F_closed}
\end{equation}
\end{proposition}
\begin{proof}
Since $\rho$ and $v$ depend only on $\theta$, the continuity equation in
the moving frame gives $(q-k)[\rho(v-u)]'=\Gamma\rho$. Along a trajectory
$\mathrm{d}F/\mathrm{d}t=-\Gamma F$ and
$\mathrm{d}\theta/\mathrm{d}t=(q-k)(v-u)$, so
$\mathrm{d}[F\rho(v-u)]/\mathrm{d}t=-\Gamma F\rho(v-u)+F(v-u)\Gamma\rho=0$.
\end{proof}
For $|c_0|\ne|c_q|$, as in the numerical cases, $\psi_J\propto c_0+c_qe^{\im\theta}$ has no nodes. The velocity is smooth on the compact ring, and the trajectories are complete. When $|c_0|=|c_q|$ a node can appear, and the qualification of Sec.~\ref{sec:particle_sector} applies.

The invariant makes the dividing mechanism transparent. If $u$ lies outside the range of $v$, the relative velocity $v-u$ never changes sign; the phase advances monotonically and $D(t)$ is periodic, with period $T=\oint\mathrm{d}\theta/(|q-k||v-u|)$. If instead $\min v<u<\max v$, fixed phases $\theta^*$ satisfy $v(\theta^*)=u$. Trajectories can then become trapped near stable phases, with $F\sim e^{\lambda t}$ and $\lambda=|q-k||v'(\theta^*)|$. The source preserves the Born weight in both cases, but the structural dynamics is entirely different.

To evaluate the distinction explicitly, take $k=0$, $u=\Omega/q$, $\phi=qx-\Omega t$ and $B(\phi)=c_0^2+c_q^2+2c_0c_q\cos\phi$. Direct calculation gives $v-u=(a\cos\phi+b)/B(\phi)$, where $a=(q-2u)c_0c_q$ and $b=qc_q^2-u(c_0^2+c_q^2)$. For $|b|>|a|$ and $a\ne0$ the untrapped period is elementary; for $a=0$ it becomes $T=2\pi(c_0^2+c_q^2)/(|q||b|)$, the corresponding limit of

\begin{equation}
T=\frac{2\pi}{|q|}\,\Bigl|\frac{2c_0c_q}{a}
+\Bigl(c_0^2+c_q^2-\frac{2c_0c_q\,b}{a}\Bigr)\frac{\operatorname{sgn}b}{\sqrt{b^2-a^2}}\Bigr|,
\label{eq:num_ring_T}
\end{equation}
and for $|b|<|a|$ the fixed phases satisfy $\cos\phi^*=-b/a$ with
\begin{equation}
\lambda=|q|\,|v'(\phi^*)|=\frac{|q\,a|\,\sin\phi^*}{B(\phi^*)}.
\label{eq:num_ring_lambda}
\end{equation}

The two dynamical regimes can be tested without changing the Born weight. Take $L=2\pi$, $k=0$, $q=2$ and $c_0=2/\sqrt5$, $c_q=\pm1/\sqrt5$. Then $N_{\mathrm B}=1$, the source coefficient follows from Eq.~\eqref{eq:ring_psi} as $j\sqrt{2\pi}=(\Omega-q^2/2)c_q$ in the scaled units, and Jensen's formula gives the initial structural departure $D(0)=\ln(1+c_q^2/c_0^2)=\ln\tfrac54$ for uniform $P_0$.

\begin{figure}[t]
\centering
\includegraphics[width=\textwidth]{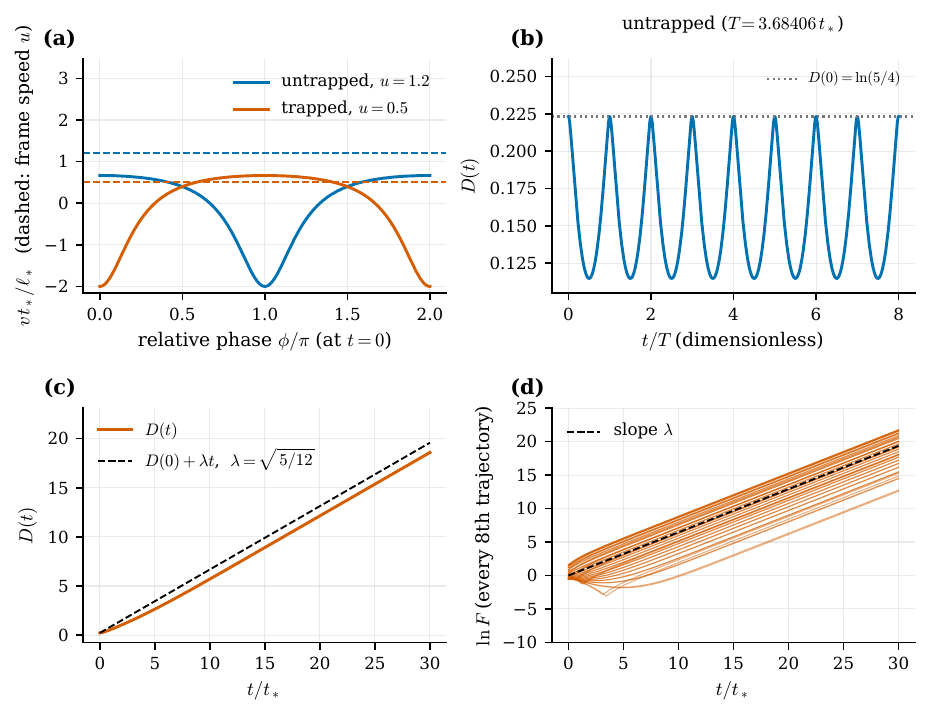}
\caption{Ring (lengths in $\ell_*=L/2\pi$, times in $t_*$). (a)~Velocity $v$ (in $\ell_*/t_*$) against the relative phase at $t=0$;
dashed lines are the frame speed $u$ (untrapped $u=1.2$, trapped
$u=0.5$). (b)~Untrapped: $D(t)$ over eight periods, $T$ from
Eq.~\eqref{eq:num_ring_T}; the dotted line is $D(0)=\ln(5/4)$.
(c)~Trapped: $D(t)$ and the line $D(0)+\lambda t$ with
$\lambda=\sqrt{5/12}$ from Eq.~\eqref{eq:num_ring_lambda}.
(d)~Trapped: $\ln F(t)$ along every eighth trajectory with the slope
$\lambda$ (dashed).}
\label{fig:ring}
\end{figure}

With $c_q>0$ and $\Omega=2.4$ ($u=1.2$), the flow is untrapped. Every phase completes a cycle with the analytical period~\eqref{eq:num_ring_T}; the numerically evaluated $D(t)$ oscillates between a minimum and its initial value (Fig.~\ref{fig:ring}b). A steady source therefore need not cause secular relaxation: this nonequilibrium ensemble returns periodically to its previous structural departure. By contrast, with $c_q<0$ and $\Omega=1$ ($u=0.5$), trajectories are trapped. Their $\ln F$ grows with the analytical rate $\lambda=\sqrt{5/12}$ of Eq.~\eqref{eq:num_ring_lambda}, and $D(t)$ grows secularly as $P$ concentrates near the stable phase while $\tilde\rho$ remains smooth (Fig.~\ref{fig:ring}c,d). The invariant, period and rate are analytical; the plotted ensemble diagnostics are computed by quadrature (\ref{app:numerics}).

Thus the same norm-preserving source mechanism supports bounded recurrence or secular structural departure, depending on the relative frame speed. Since $D(0)>0$, the ring describes the evolution of existing nonequilibrium, not its creation from an equilibrated state. We next ask what a source does when both channels are available and the initial ensemble is in equilibrium.

\subsection{Harmonic Oscillator Examples}
\label{sec:ho}
The oscillator brings a new ingredient: the guiding field can develop nodes near the moving ensemble while both nonequilibrium channels respond. Its eigenenergies are $E_n=\hbar\omega(n+\tfrac12)$. An additive drive resonant with $\varphi_n$ has frequency $\Omega=E_n/\hbar$, not a transition frequency; unlike a multiplicative perturbation such as $Fx$, it breaks $U(1)$ phase symmetry and does not merely couple adjacent energy levels.

\exhead{HO.1}{resonant drive of an equilibrated ensemble}The central question is whether a source can produce a structural departure when none existed initially. Set $\psi_0=\varphi_0$ and $P_0=\varphi_0^2$, so $f_0=1$, and drive the first excited mode with $J=j_0e^{-\im\Omega t}\varphi_1(x)$ at $\Omega=3\omega/2$.\label{sec:ho1} The solution remains in the span of $\varphi_0$ and $\varphi_1$: $\psi_J=c_0(t)\varphi_0+c_1(t)\varphi_1$, where

\begin{equation}
c_0=e^{-\im\omega t/2},
\qquad
c_1=-\im\frac{j_0 t}{\hbar}e^{-\im 3\omega t/2}.
\label{eq:ho1_coefficients}
\end{equation}
The linear growth of $c_1$ is the resonance: the source pumps the first excited state at a constant rate. With $\varphi_1=\sqrt2\,a\,x\,\varphi_0$ and $a=\sqrt{m\omega/\hbar}$, the velocity field is
\begin{equation}
v(x,t) = \frac{\hbar}{m}\operatorname{Im}\!\left[
\frac{\sqrt{2}\,a\,c_1}{c_0 + \sqrt{2}\,a\,c_1\,x}
\right].
\label{eq:ho1_v}
\end{equation}
The source rate $\Gamma=(2/\hbar)\operatorname{Im}(J/\psi_J)$ is spatially varying, so both diagnostics can evolve. To compute the structural response we integrate $q$ and $\ln F$ together using Eq.~\eqref{eq:main_ode}. The challenge is near the zeros of the field. The denominator in Eq.~\eqref{eq:ho1_v} vanishes at $x^*(t)=-c_0/(\sqrt2\,a\,c_1)=-\im\hbar e^{\im\omega t}/(\sqrt2\,a\,j_0t)$, which crosses the real axis at $t=(n+\tfrac12)\pi/\omega$. The velocity is singular at an exact node, but trajectories passing nearby can acquire large finite velocities. A crossing of the complex-field zero through the real axis does not establish that any trajectory hits it; almost-everywhere completeness remains conditional on Assumption~\ref{ass:H}.

This local effect has a global consequence. At late times the normalized field tends to $\tilde\rho\to\varphi_1^2$, while at fixed $x\ne0$ its guidance velocity is $O(\cos(\omega t)/t)$. That estimate is not uniform: $x^*(t)$ approaches the origin as $1/t$, and $v(0,t)$ grows like $t\cos\omega t$. It suggests, but does not prove, limited net displacement for typical trajectories. The computed ensemble nevertheless develops an apparent nonzero plateau in $D(t)$ over the finite horizons studied, rather than returning to its initial equilibrium.

How large is this change compared with a distribution that does not move? Frozen particles would give $D_{\mathrm{KL}}(\varphi_0^2\|\varphi_1^2)=\gamma_E+\ln2\approx1.27$. The observed departure is much smaller because particles follow the changing guidance field. We evaluate $D$ with the trajectory-based Lagrangian average of Eq.~\eqref{eq:D_lagrangian}; the adaptive treatment of near-nodal passages and the ensemble quadrature are documented in \ref{app:numerics}.

\begin{figure}[t]
\centering
\includegraphics[width=\textwidth]{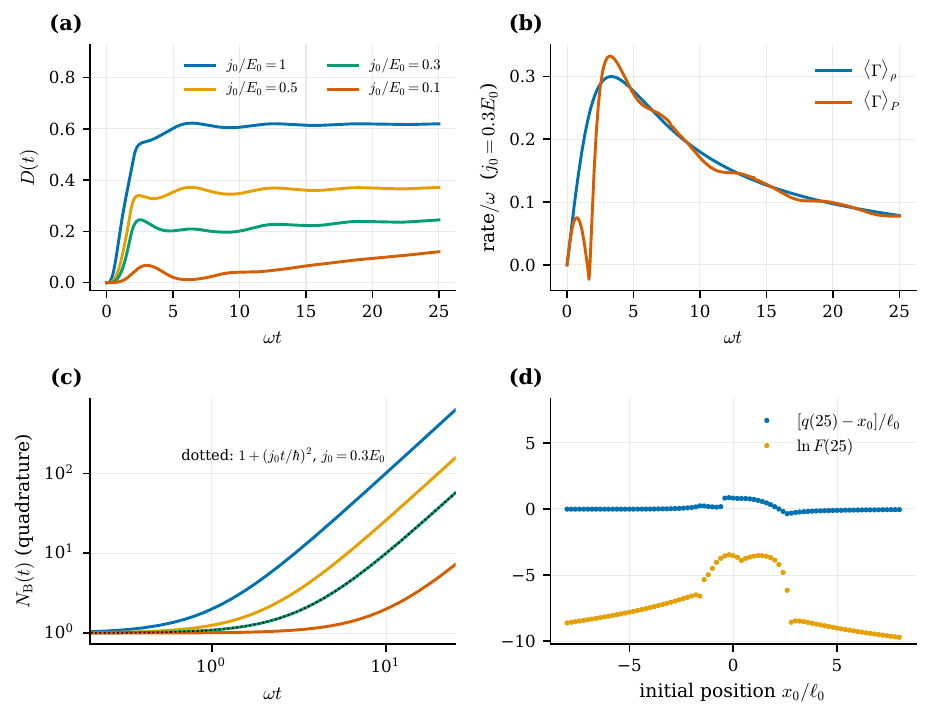}
\caption{HO.1 (scaled oscillator units $\ell_0$, $\omega^{-1}$, $E_0$). (a)~$D(t)$ for $j_0/E_0=1,0.5,0.3,0.1$. (b)~The two
averages $\langle\Gamma\rangle_\rho$ and $\langle\Gamma\rangle_P$ for $j_0/E_0=0.3$, with rates in units of $\omega$. (c)~$N_{\mathrm B}(t)$ obtained by quadrature of $|\psi_J|^2$
(solid, all four $j_0$) and $1+(j_0t/\hbar)^2$ for $j_0/E_0=0.3$ (dotted).
(d)~Displacement $[q(t)-x_0]/\ell_0$ and dimensionless $\ln F(t)$ at $\omega t=25$ against $x_0/\ell_0$,
$j_0/E_0=0.3$ (every 40th node).}
\label{fig:ho1}
\end{figure}

For $j_0$ between $0.1$ and $1$, the curves rise rapidly around the nodal encounters and then level off (Fig.~\ref{fig:ho1}a). The difference $\langle\Gamma\rangle_\rho-\langle\Gamma\rangle_P$, rather than either average separately, controls this response (Fig.~\ref{fig:ho1}b). The particle motion is concentrated in narrow regions of the initial ensemble: most particles move little, whereas near-nodal trajectories undergo localized displacement and sharp changes of $\ln F$ (Fig.~\ref{fig:ho1}d). The persistence of $D$ is thus tied to localized transport, not to a bulk shift of the Gaussian.

\begin{figure}[t]
\centering
\includegraphics[width=0.78\textwidth]{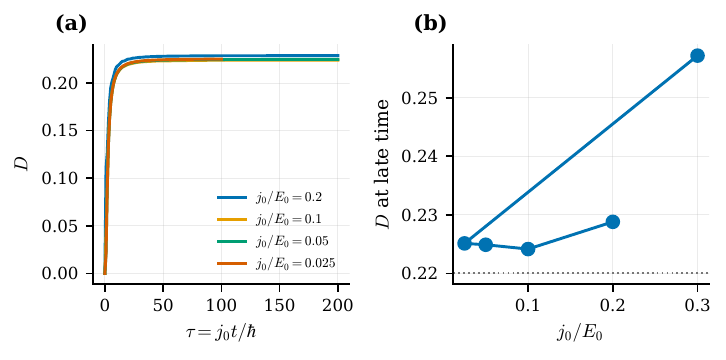}
\caption{Weak-drive limit of HO.1. (a)~$D$ against $\tau=j_0t/\hbar$ for
$j_0/E_0=0.2,0.1,0.05,0.025$, computed to $\tau=200,200,200,100$. (b)~Value at the last time computed against $j_0/E_0$
(the point at $j_0/E_0=0.3$ is HO.4 with $c=1$, at $\omega t=600$).}
\label{fig:weak}
\end{figure}

Reducing the drive does not remove the response over the rescaled observation windows studied. Because appreciable displacements accumulate on a time scale $\hbar/j_0$, curves for $j_0/E_0=0.2,0.1,0.05,0.025$ nearly collapse when plotted against $\tau=j_0t/\hbar$ (Fig.~\ref{fig:weak}a). Their endpoint values approach $D\approx0.22$ as the drive is decreased over the stated finite horizons, a numerical observation rather than an established weak-drive or infinite-time limit (\ref{app:numerics}). The plateau remains far below the frozen-particle value. Larger drives produce larger late-horizon departures, a dependence examined again in HO.4.

\exhead{HO.2}{amplitude-only drive}HO.1 shows that a nonuniform source rate can reshape an initially equilibrated ensemble. The contrasting question is whether a source can change the Born weight without changing the structure at all. Choose $\psi_0=\varphi_0$ and $J=j(t)\varphi_0(x)$, where $j(t)$ is a prescribed scalar.\label{sec:ho2} The field stays in one mode: $\psi_J=c(t)\varphi_0$, with $\im\hbar\dot c=E_0c+j(t)$. The rate $\Gamma=(2/\hbar)\operatorname{Im}(j/c)$ is spatially uniform, so $\hat\Gamma=0$ and $D$ is conserved.

\begin{figure}[t]
\centering
\includegraphics[width=0.66\textwidth,trim=0 0 150 0,clip]{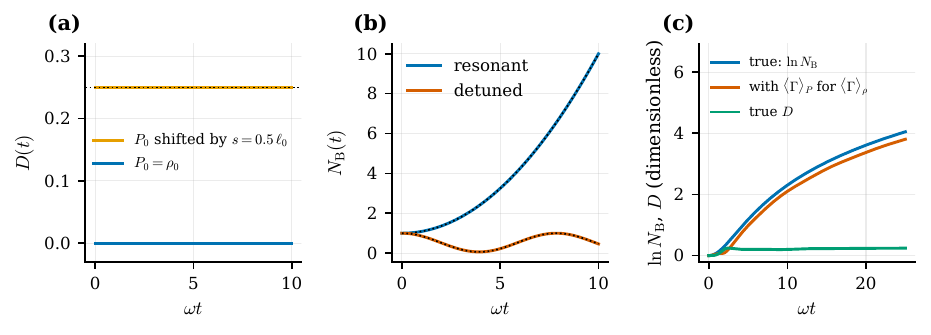}
\caption{HO.2 (scaled oscillator units, shift $s$ in $\ell_0$). (a)~$D(t)$ for $P_0=\rho_0$ ($D\equiv0$) and for $P_0$
shifted by $s=0.5$ (constant $D=s^2$, dotted). (b)~$N_{\mathrm B}(t)$ by
quadrature (solid) and $|c(t)|^2$ (dotted) for resonant and detuned
drive.}
\label{fig:ho2}
\end{figure}

For the equilibrated choice $P_0=\varphi_0^2$ we obtain $D\equiv0$ while $N_{\mathrm B}(t)=|c(t)|^2/|c(0)|^2$ changes. The velocity $v=(\hbar/m)\operatorname{Im}(\partial_x\varphi_0/\varphi_0)$ is independent of $c(t)$; thus the particles remain stationary as in the source-free ground state. We exclude source histories at which $c(t)$ vanishes. The result is not restricted to initially equilibrated particles: shifting $P_0$ by $s=0.5$ gives $D(0)=s^2=0.25$, and this value remains fixed (Fig.~\ref{fig:ho2}). A uniform $\Gamma$ cannot drive the structural channel, whether or not its initial value is zero. Spatial variation of $\Gamma$ is necessary for structural evolution, though the ring shows it is not sufficient for growth.

This solution has the form of a state-proportional drive, $J=(j/c)\psi_J$ (Remark~\ref{rem:exact_rescaling_2}), but the source itself is specified exogenously. The proportionality occurs only because the source and field occupy the same mode. HO.2 therefore isolates a special amplitude-only response, not a general property of exogenous sources. In HO.1 the first-excited-mode component instead makes $\Gamma$ nonuniform and activates both channels. This comparison also explains why the particle-weighted rate alone cannot determine $\ln N_{\mathrm B}$: the difference between that average and the field-weighted average is precisely what changes $D$.

\exhead{HO.3}{coherent-state packet}A coherent packet oscillates through the spatial region occupied by the driven mode. Might that motion delay the influence of the source on its guided particles? We test this by replacing the oscillator ground state with the coherent state $|\alpha\rangle$, where $\alpha$ is dimensionless, while retaining $J=j_0e^{-\im\Omega t}\varphi_1$ at $\Omega=3\omega/2$ and $P_0=\rho_0$.\label{sec:ho3} The field is $\psi_J=e^{-\im\omega t/2}|\alpha e^{-\im\omega t}\rangle+c_1(t)\varphi_1$. Its zero-point phase is essential to interference with the driven mode: the coherent packet moves rigidly at frequency $\omega$, while $c_1$ grows as in HO.1.

The centre follows $x_c(t)=\sqrt2\operatorname{Re}(\alpha e^{-\im\omega t})/a$. At its turning points the driven-to-coherent amplitude ratio is $(j_0t/\hbar)\,2|\alpha|e^{-|\alpha|^2}$; setting this ratio to unity suggests an exponentially large time,

\begin{equation}
t_\times\sim\frac{\hbar}{2j_0|\alpha|}\,e^{|\alpha|^2}.
\label{eq:ho3_crossover}
\end{equation}
But this estimate probes only the packet's turning points. Twice each period it crosses the region $|x|\lesssim a^{-1}$, where $\varphi_1$ is not exponentially suppressed. A turning-point amplitude ratio therefore need not measure the time at which a trajectory first responds. We test that distinction directly, integrating the packet-centre trajectory rather than inferring its behavior from the field amplitude alone (\ref{app:numerics}).

\begin{figure}[t]
\centering
\includegraphics[width=\textwidth]{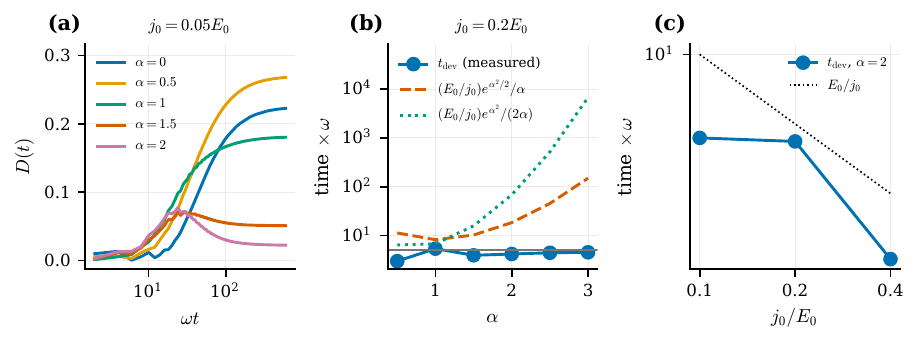}
\caption{HO.3 (scaled oscillator units; times in $\omega^{-1}$). (a)~$D(t)$ for $\alpha=0,\dots,2$, $j_0=0.05E_0$.
(b)~Scaled departure time $\omega t_{\rm dev}$ of the packet-centre trajectory, first defined by $|q-q_{\rm free}|/\ell_0>0.3$, against dimensionless $\alpha$, with the guide $(E_0/j_0)e^{\alpha^2/2}/\alpha$ (dashed),
$\omega t_\times\sim(E_0/j_0)e^{\alpha^2}/(2|\alpha|)$ from Eq.~\eqref{eq:ho3_crossover} (dotted), and $E_0/j_0$ (grey).
(c)~$\omega t_{\rm dev}$ against $j_0/E_0$ at $\alpha=2$.}
\label{fig:ho3}
\end{figure}

Without the drive, a particle starting at $x_0=\sqrt2\alpha/a$ would follow $q_{\rm free}(t)=(\sqrt2\alpha/a)\cos(\omega t)$. In the driven problem it departs from this rigid motion already during the first passage near the origin. Defining the numerical departure time by $|q(t)-q_{\rm free}(t)|/\ell_0>0.3$, the response time has neither the exponential dependence on $\alpha$ suggested by Eq.~\eqref{eq:ho3_crossover} nor a simple $1/j_0$ scaling (Fig.~\ref{fig:ho3}b,c); at $\alpha=3$ the turning-point estimates lie one to three orders of magnitude later. Coherent motion therefore does not screen the earliest trajectory response in this test.

It does alter the structural evolution. The finite-horizon value of $D$ depends on $\alpha$ nonmonotonically and is small for $\alpha\gtrsim1.5$; for these displacements $D(t)$ can rise and then relax (Fig.~\ref{fig:ho3}a). Coherent preparation changes how the driven mode and transported ensemble meet, but does not supply a universal delay mechanism or a universal late-time structural value.

\exhead{HO.4}{fixed-source rescaling}The preceding examples varied the source and the initial wavefunction's shape. A different question arises from scale covariance: does the initial amplitude of the guiding wave matter if the exogenous source is held fixed? We return to HO.1, take $\psi_0=c\varphi_0$ with a dimensionless factor $c>0$, keep $J=j_0e^{-\im\Omega t}\varphi_1$ at $\Omega=3\omega/2$, and choose $P_0=\varphi_0^2$.\label{sec:feedback} For every $c$, $\tilde\rho_0=\varphi_0^2$ and $D(0)=0$, but $N_{\mathrm B}(0)=c^2$. The exact field is $\psi_J=c\,e^{-\im\omega t/2}\varphi_0-\im(j_0t/\hbar)e^{-\im3\omega t/2}\varphi_1$, with $N_{\mathrm B}(t)=c^2+(j_0t/\hbar)^2$. The guidance velocity and $\Gamma$ depend on the ratio $j_0/c$: strengthening the initial wave relative to a fixed source weakens the relative drive.

This is a dynamical distinction, not merely a change of normalization. The criterion of Proposition~\ref{prop:scale_covariance} can be evaluated directly. With $\psi_{\mathrm{inh}}=c_1\varphi_1$, the $c=1$ case has $r=\psi_{\mathrm{inh}}/\psi_J=\zeta x/(1+\zeta x)$, where $\zeta=\sqrt2\,a\,c_1/c_0=-\im\sqrt2\,a\,(j_0t/\hbar)e^{-\im\omega t}$. For real $x$, $\operatorname{Im}r=x\operatorname{Im}\zeta/|1+\zeta x|^2$ and $\operatorname{Im}\zeta=-\sqrt2\,a\,(j_0t/\hbar)\cos\omega t$. Except at the specified discrete node-crossing times $t=(n+\tfrac12)\pi/\omega$, this is nonzero when $x\ne0$. Thus $r$ is neither real nor spatially constant, the velocity-invariance condition fails, and varying $c$ at fixed $J$ changes the guidance.

At short times $v(x,t)=-\sqrt2\,a\,j_0t/(mc)+O(t^2/c^2)$ is spatially uniform to leading order. A larger $c$ therefore slows the early motion and keeps the transported ensemble closer to $\varphi_0^2$, delaying the development of $D$ from its common initial value zero. The initial amplitude $N_{\mathrm B}(0)$ consequently controls the structural response through the relative source strength. Jointly rescaling $c$ and $j_0$, by contrast, leaves $D(t)$ invariant (Remark~\ref{rem:joint_rescaling}): the pair $(c,j_0)$ is structurally equivalent to $(1,j_0/c)$. This is the distinction between varying a field against a fixed reference and rescaling the entire field--source configuration.

For fixed time, $c\to\infty$ gives $v\to0$ and $D\to0$. It does not follow that an arbitrarily weak relative source has no effect at long times: the driven mode becomes comparable to the initial field on the scale $\hbar c/j_0$. In the finite integrations, the late-horizon values approach the weak-drive plateau $\approx0.22$ found in HO.1, suggesting but not proving that the large-$c$ and long-time limits need not commute.

\begin{figure}[t]
\centering
\includegraphics[width=\textwidth]{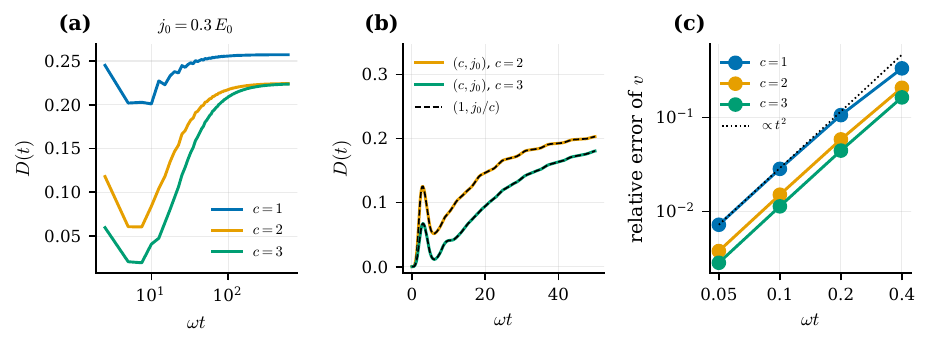}
\caption{HO.4, $j_0=0.3E_0$ (scaled oscillator units). (a)~$D(t)$ for $c=1,2,3$. (b)~Joint
rescaling: $D$ for $(c,j_0)$ (colours) and $(1,j_0/c)$ (dashed) coincide.
(c)~Dimensionless relative error of the leading-order velocity on $|x|/\ell_0\le2$.}
\label{fig:ho4}
\end{figure}

For $j_0=0.3$ and $c=1,2,3$, the computed ordering is $D^{(1)}(t)>D^{(2)}(t)>D^{(3)}(t)>0$ at output times after the common start (Fig.~\ref{fig:ho4}a). The curves move closer over the observed window without establishing a shared limit. Their late-horizon values are consistent with a dependence on $g=j_0/c$, nearly flat around $0.22$ for $g\lesssim0.15$ but larger at $g=0.3$ (Fig.~\ref{fig:weak}b). The exact covariance of $(c,j_0)$ and $(1,j_0/c)$ is reproduced numerically (Fig.~\ref{fig:ho4}b), as is the short-time velocity approximation (Fig.~\ref{fig:ho4}c; numerical details in \ref{app:numerics}). The algebraic symmetry is exact; the behavior of $D$ over these finite horizons is numerical.

\subsection{Two-Body Dynamics Under Source: Minimal Model}
\label{sec:twobody}
\exhead{2B}{coupled oscillators, nonlocal guidance and entanglement}The final model asks what is added when the guidance problem has two particles. A source in configuration space can make the velocity of one particle depend on the position of the other, even when the field equation separates after a change of coordinates. The two-body ISE contains $J(x_1,x_2,t)$ in addition to the usual kinetic and potential terms, with guidance $\dot x_k=(\hbar/m_k)\operatorname{Im}(\partial_{x_k}\psi_J/\psi_J)$ for $(x_1,x_2)\in\mathbb R^2$. The diagnostics $\tilde\rho$ and $D$ are now integrated over this two-particle space.

We choose two oscillators of frequency $\omega$, coupled by $\tfrac12k(x_1-x_2)^2$. The normal coordinates $X=(x_1+x_2)/\sqrt2$ and $x=(x_1-x_2)/\sqrt2$ separate the Hamiltonian into frequencies $\omega_X=\omega$ and $\omega_x=\sqrt{\omega^2+2k/m}$. Drive the relative mode with $J=j_0e^{-\im\Omega t}\varphi_1^{(x)}(x)\varphi_0^{(X)}(X)$, where $\Omega=E_{01}/\hbar=\tfrac12(\omega+3\omega_x)$. This is resonant with the complete state $\varphi_0^{(X)}\varphi_1^{(x)}$: the centre-of-mass zero-point energy must be included, since $3\omega_x/2$ alone would be detuned by $\omega/2$. The centre-of-mass mode remains in its ground state; after removing a common phase, the relative part has the HO.1 coefficients~\eqref{eq:ho1_coefficients} with $\omega\to\omega_x$.

\begin{figure}[t]
\centering
\includegraphics[width=\textwidth]{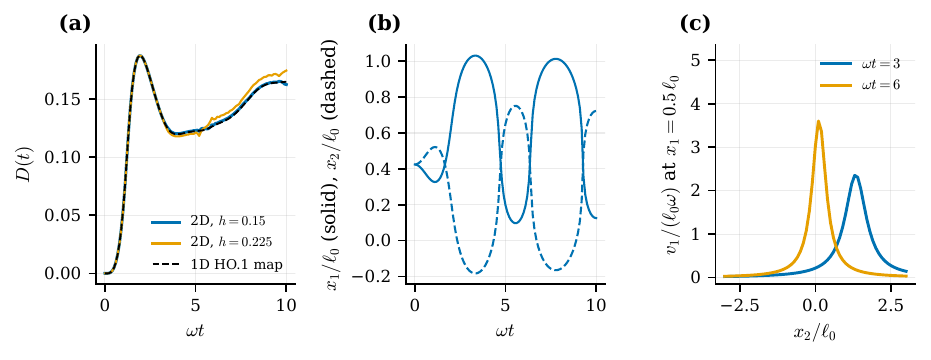}
\caption{Example 2B ($k/m\omega^2=0.5$, $j_0=0.3E_0$; lengths in $\ell_0$, times in $\omega^{-1}$, velocities in $\ell_0\omega$). (a)~$D(t)$ on two tensor grids (solid; relative-coordinate spacings $h/\ell_0=0.15,0.225$) and
the converged one-dimensional reduction (dashed). (b)~Coordinates of five trajectories started
at $X/\ell_0=0.6$ ($x_1$ solid, $x_2$ dashed). (c)~$v_1$ at fixed $x_1/\ell_0=0.5$ as a function of $x_2$
at $t=3,6$.}
\label{fig:twobody}
\end{figure}

The separation simplifies the field calculation, but not the physical particle partition. From the chain rule $\dot x_1=(\dot X+\dot x)/\sqrt2$ and $\dot x_2=(\dot X-\dot x)/\sqrt2$. The relative velocity depends on $x=(x_1-x_2)/\sqrt2$, so $v_1$ depends on both particle coordinates, as seen directly in Fig.~\ref{fig:twobody}c. Thus a source confined to a normal mode yields nonlocal Bohmian guidance in particle coordinates. For the numerical characteristics we integrate Eq.~\eqref{eq:main_ode} in $(x_1,x_2,\ln F)$ over finite intervals. A nodal line $x=x^*$ occurs at the discrete times $t^*$ when $\chi(x^*,t^*)=0$; we integrate sampled trajectories away from exact nodal encounters, while almost-everywhere completeness in the continuum description remains the hypothesis of Assumption~\ref{ass:H}.\label{sec:twobody_example}

For the calculation take $m_1=m_2=\hbar=\omega=1$, $k=0.5$ ($\omega_x=\nv{twoW}$), $j_0=\nv{twoJ}$, $P_0=\rho_0$ and $t\le\nv{twoTend}$. The normal-mode reduction provides an independent test: each two-dimensional trajectory must agree with HO.1 after $j_0\to j_0/\omega_x$, $t\to\omega_xt$ and $x\to\sqrt{\omega_x}x$. It does, with $\dot X=0$, $v_1+v_2=0$, and agreement of both computational routes for $D$. The structural diagnostic rises to $0.187$ and subsequently lies near $0.165$ (Fig.~\ref{fig:twobody}a).

The final value deserves a numerical qualification. Because $\psi_J=\varphi_0^{(X)}(X)\chi(x,t)$ and the centre-of-mass factor is static, the initial ensemble average reduces exactly to one dimension. The quoted $D(t)$ therefore comes from the converged relative-coordinate quadrature. Direct two-dimensional integration confirms the pointwise reduction, but does not resolve $D$ at the $10^{-3}$ level on the grids employed (\ref{app:numerics}, Table~\ref{tab:twobody_refine}). The source-driven structural response is real; the reduced calculation, rather than a coarse tensor grid, supplies its stated numerical accuracy.

Normal-mode factorization does not imply particle separability. The two-particle wavefunction is generally entangled in $(x_1,x_2)$: with unequal normal-mode frequencies the initial pure state already has reduced-state purity $2\sqrt{\omega_x}/(1+\omega_x)\approx0.985$. The driven purities are approximately $0.75$ at $t=5$ and $0.58$ at $t=10$, from a Schmidt decomposition on a $241\times241$ grid. Since a normalized pure bipartite state's purity below one signals entanglement, the drive increases pre-existing entanglement; it does not create entanglement from an initially product state.

Three properties must remain distinct. Entanglement concerns the particle partition of the quantum state; nonlocal guidance concerns the dependence of $v_1$ on $x_2$ through the phase (a real entangled state can have zero instantaneous velocity); and the reduction of $D$ concerns a special separability of the normal-coordinate ensemble calculation. The last does not eliminate either of the first two. Systematic entanglement dynamics, marginal distributions and nonseparable sources lie beyond this minimal model.

The examples now give a physical reading of the two-channel decomposition rather than a single pattern of departure from equilibrium. Infinite space first exposes its domain of validity: idealized point and plane-wave fields can guide locally while lacking normalized global diagnostics, with the off-resonant zero-field exceptions noted above. The regularized impulse supplies an exact jump rule confirmed by finite-pulse transport; F$\infty$.3 connects this to sustained forcing without a computed global $D$. The ring then makes the distinction between norm and structure particularly sharp: the Born weight is fixed, but an existing structural departure can recur periodically or grow secularly.

The oscillators show what changes when both channels are available. HO.1 produces a structural departure from initial equilibrium through localized near-nodal transport, far below the frozen-particle value over the finite horizons computed. HO.2 isolates the amplitude-only response. HO.3 shows that coherent initial motion alters the response without providing the anticipated exponential screening, and HO.4 demonstrates why initial field amplitude measured relative to a fixed source affects the trajectories. Finally, the coupled two-body model brings nonlocal guidance and evolving particle entanglement into the same sourced dynamics, while retaining an exact reduction of the structural diagnostic.

Table~\ref{tab:summary} separates what is analytical, what is numerically established and what has not been computed. Across these different regimes one principle remains: $\mathrm{d}D/\mathrm{d}t$ is determined by $\langle\Gamma\rangle_\rho-\langle\Gamma\rangle_P$, not by the sign of $\Gamma$ alone. Injecting Born density can accompany structural relaxation, and extracting it can accompany growing structural departure. The source acts not only on the field's norm but on the way its changing structure guides and redistributes the particles.

\begin{table}[t]
\centering
\footnotesize
\caption{Summary of examples by channel and provenance. The columns list the configuration-space domain, whether $N_{\mathrm{B}}$ is finite, the behavior of the structural diagnostic $D$, the behavior of the amplitude diagnostic $\ln N_{\mathrm{B}}$, and what the example establishes together with its analytical or numerical provenance. N/A means that the global diagnostic is undefined, and ``not computed'' that it was not evaluated; neither means zero.}
\label{tab:summary}
\newcolumntype{R}[1]{>{\raggedright\arraybackslash}p{#1}}
\setlength{\tabcolsep}{3pt}
\begin{tabular}{R{1.7cm} R{0.95cm} R{1.0cm} R{2.0cm} R{1.45cm} R{4.7cm}}
\toprule
Example & Domain & $N_{\mathrm{B}}$ finite? & Structural $D$ &
Amplitude $\ln N_{\mathrm{B}}$ & Established; provenance \\
\midrule
F$\infty$.1 point & $\mathbb{R}^3$ & No & N/A & N/A & Local guidance only; fully analytic \\
F$\infty$.2 reg.\ imp.\ & $\mathbb{R}$ & Yes & Jump rule & Active & Ratio field jumps across a sudden kick; analytic rule, numerically validated pulse transport \\
F$\infty$.3 pulse & $\mathbb{R}^3$ & Yes ($g\in L^2$) & Not computed & Active & Interpolates between impulse and drive; model only \\
Ring untrapped & $S^1$ & Yes & Periodic & Constant & Pre-existing $D(0)>0$ evolves periodically; analytic invariant and period, numerical ensemble $D(t)$ \\
Ring trapped & $S^1$ & Yes & Secular, rate $\lambda$ & Constant & Concentration of $P$ at fixed Born weight; analytic criterion and rate, numerical $D(t)$ \\
HO.1 ($\varphi_1$) & $\mathbb{R}$ & Yes & $\approx0.22$ (weak drive, finite horizon) & Active & $D$ generated from equilibrium, far below frozen value; analytic field, numerical transport \\
HO.2 ($\varphi_0$) & $\mathbb{R}$ & Yes & Unchanged ($D=0$ here) & Active & Uniform $\Gamma$ drives $N_{\mathrm B}$ only; fully analytic \\
HO.3 (coherent) & $\mathbb{R}$ & Yes & $\alpha$-dependent & Active & Coherent state alters response; analytic field, numerical transport \\
HO.4 (fixed-source scale) & $\mathbb{R}$ & Yes & $c$-dependent & Active & Field amplitude relative to fixed source matters; exact identity, numerical $D(t)$ \\
2B (coupled osc.) & $\mathbb{R}^2$ & Yes & Reduces to 1D & Active & Nonlocal guidance, particle entanglement; analytic field, numerical transport \\
\bottomrule
\end{tabular}
\end{table}

\section{Concluding Remarks}
\label{sec:conclusion}
The source term $J$ makes the guiding-field density $\rho=|\psi_J|^2$ non-conserved. What would be an obstruction in orthodox quantum mechanics is not one in the de~Broglie--Bohm ontology of Ref.~\cite{mikki2026AdP}: probability is carried by the particle distribution $P$, while $\rho$ belongs to the field that guides it. The question is therefore not how to restore conservation of $\rho$, but how to define equilibrium when the two densities can evolve differently. We have shown that the departure has two independent components. The structural diagnostic $D=D_{\mathrm{KL}}(P\,\|\,\tilde\rho)$ compares their normalized shapes, and the amplitude diagnostic $N_{\mathrm B}$ measures the field's Born weight. The Born rule holds if and only if $D=0$ and $N_{\mathrm B}=1$ (Proposition~\ref{prop:born_two_conditions}). Both rates derive from $J/\psi_J$, but $D$ responds to the difference $\langle\Gamma\rangle_\rho-\langle\Gamma\rangle_P$, not simply to the sign of the local source rate.

The distinction also gives amplitude a dynamical role. Proposition~\ref{prop:scale_covariance} shows that a linear endogenous generator retains rescaling symmetry, whereas an exogenous source supplies an independent reference: the strength of the field relative to that source can change the guidance and the structural response. The physical status of absolute field amplitude remains a premise of the adopted ontology, not an inference from single-time particle-position statistics, which are normalized by construction. A calibrated source together with the normalized state and particle distribution could, in principle, reconstruct both diagnostics (\ref{app:measure}); this conditional possibility is neither needed for the present results nor a substitute for a general theory of subquantum measurement.

The driven examples answer the dynamical question more concretely. The free-space impulse reveals the limits of global equilibrium diagnostics, while a regularized kick gives an exact jump in the ratio field. On the ring the Born weight remains fixed, yet the structural departure can recur periodically or grow secularly according to the trapping regime. The oscillator models show the complementary possibilities: a resonant source can generate structural nonequilibrium from an initially equilibrated ensemble through localized, near-nodal transport, whereas a drive in the occupied mode changes the amplitude alone. Coherent preparation and initial field amplitude relative to a fixed source further alter the response. The coupled two-body oscillator adds nonlocal guidance and evolving particle entanglement without changing the formal two-channel identities.

The numerical work is part of these conclusions, not merely a check of the algebra. Exact or semi-analytical guiding fields are paired with source-aware particle transport and independently evaluated ensemble diagnostics; the resulting regimes cannot be read off from the equilibrium classification alone. The oscillator plateau and related weak-drive behavior are observations over long but finite numerical horizons. Ensemble results remain conditional on Assumptions~\ref{ass:A}--\ref{ass:E}, including the no-leakage and trajectory-completeness hypotheses; no general theorem of global existence is claimed. In the source-free, initially equilibrated sector, $D(0)=0$ and $N_{\mathrm B}(0)=1$ give standard quantum mechanics, while source-free nonequilibrium preserves its initial diagnostic values. The driven regimes illustrate the additional dynamics accessible in the source-extended theory and motivate further study of nonseparable sources, many-body ensembles and entanglement evolution.

\appendix
\section{Mathematical Preliminaries}
\label{app:math}\label{rem:duhamel_wellposed}
The continuum formulation uses Hilbert- and Banach-space results to distinguish existence of the guiding field from existence of particle trajectories. We collect the conventions and sufficient conditions needed for the former; the latter remains subject to Assumptions~\ref{ass:A} and~\ref{ass:H}.

Set $\mathcal H=L^2(\mathcal C)$ and let $X$ be a Banach space. On an interval $I$, $C^k(I;X)$ consists of $k$-times continuously differentiable $X$-valued functions, while $L^p(I;X)$ consists of measurable functions with $\int_I\|f(t)\|_X^p\,\mathrm dt<\infty$. For $p=1$ the integral is a Bochner integral and satisfies $\|\int f\|_X\le\int\|f\|_X$. We write $L^1_{\mathrm{loc}}$ for integrability on bounded subintervals, $\mathcal D(H^k)$ for the domain of $H^k$ (including the requisite lower powers), and $\mathcal S'$ for tempered distributions, including the Dirac delta.

A \emph{mild solution} is represented by the Duhamel formula~\eqref{eq:duhamel}; a \emph{classical solution} has an $\mathcal H$-valued time derivative and satisfies the ISE pointwise in time. For $H$ self-adjoint with domain $\mathcal D(H)$, the following are standard sufficient conditions \cite{cazenave2003semilinear,pazy1983semigroups}:

\begin{enumerate}
\item If $\psi_0\in\mathcal{H}$ and $J\in L^1([0,T];\mathcal{H})$ for every $T>0$, then $\psi_J\in C([0,\infty);\mathcal{H})$ is the unique mild solution of the ISE~\eqref{eq:ise}.
\item If in addition $\psi_0\in\mathcal{D}(H^k)$ and $J\in C([0,T];\mathcal{H})\cap L^1([0,T];\mathcal{D}(H^k))$ for sufficiently large $k$, then $\psi_J\in C^1([0,T];\mathcal{H})\cap C([0,T];\mathcal{D}(H^k))$ is a classical solution. For Schr\"odinger operators with smooth potentials, local elliptic regularity gives $\mathcal{D}(H^k)\subset H^{2k}_{\mathrm{loc}}$, so that for $k$ large enough the Sobolev embedding makes $\psi_J$ of class $C^2$ in space, which is the regularity required by Assumption~\ref{ass:A}; for a general self-adjoint $H$ this spatial regularity is an additional hypothesis.
\item If $J$ is singular in time only, for instance $J=j(x)\,\delta(t-t_0)$ with $j\in\mathcal{H}$, the Duhamel integral is still well defined and gives $\psi_J(t)=U(t,0)\psi_0-(\im/\hbar)U(t,t_0)\,j\in\mathcal{H}$ for $t>t_0$, so that $N_{\mathrm{B}}$ is finite; a temporal singularity does not by itself make the Born weight diverge. If instead the spatial profile of $J$ is singular or does not decay (a point source, a plane wave on an unbounded domain), the Duhamel integral is interpreted distributionally and $\psi_J\in\mathcal{S}'$ rather than $L^2$; then the Born weight is infinite whenever the resulting plane-wave or point-source profile is nonzero; at any exceptional time when the resulting field vanishes identically its Born weight is zero instead. In either case the normalized quantities are undefined (Remark~\ref{rem:NB_square_integrability}), and only local kinematic quantities are analysed.
\end{enumerate}

For the guidance equation, the corresponding question is local uniqueness of paths. Continuity in time and local Lipschitz regularity in space of the velocity field, as required in Assumption~\ref{ass:A}, provide it. In the source-free case, Lemma~6.1 of Ref.~\cite{berndl1995global_existence} supplies smooth classical solutions under smooth potentials and initial states that are $C^\infty$-vectors of $H$ (in every power-domain); explicit sourced fields can be checked directly. These are local regularity statements, not a proof of global trajectory existence. Almost-everywhere completeness is assumed when needed through Assumption~\ref{ass:H}. Sufficient conditions are investigated separately in Ref.~\cite{mikki2026existence}, whose results are not used here.

\section{Proof of Proposition~\ref{prop:scale_covariance}}
\label{app:scale_proof}
The scale-covariance statements separate joint rescaling from rescaling at fixed source. The proof first establishes the preserved quantities for a linear generator, then derives the affine change at fixed exogenous source, and finally characterizes when that change leaves the guidance velocity invariant. No additional dynamical assumptions enter these algebraic steps.
\begin{enumerate}
\item If $\mathcal{K}$ is $\mathbb{C}$-linear,
$\im\hbar\partial_t(c\psi_J) = H(c\psi_J) + c\mathcal{K}[\psi_J]
= H(c\psi_J) + \mathcal{K}[c\psi_J]$, so $c\psi_J$ solves the same
ISE. The velocity $v^{\psi_J} = (\hbar/m)\operatorname{Im}(\nabla\psi_J/\psi_J)$
is unchanged by $\psi_J\to c\psi_J$. Next,
$W = \mathcal{K}[\psi_J]/\psi_J = c\mathcal{K}[\psi_J]/(c\psi_J)$ is
unchanged, and so is $\Gamma$. Further,
$\tilde\rho = |\psi_J|^2/N_{\mathrm{B}}$ is unchanged because numerator
and denominator both scale by $|c|^2$, and $D$ depends only on $P$ and
$\tilde\rho$. The positively homogeneous case repeats the argument for
$c>0$.
\item Since $\mathcal{K}$ is linear, the solution map is affine in
the data $(\psi_0,J_0)$ and linear in $\psi_0$ at fixed $J_0$:
$\psi_J = U\psi_0 + \psi_{\mathrm{inh}}$, where $U\psi_0$ is the
solution with $J_0$ switched off and $\psi_{\mathrm{inh}}$ the solution
with zero initial data. Hence
$\psi_J^{(c)} = cU\psi_0 + \psi_{\mathrm{inh}}
= c\psi_J + (1-c)\psi_{\mathrm{inh}}$, and dividing by $\psi_J$ gives
the factor $c+(1-c)r$.
\item From item~(ii),
$\nabla\psi_J^{(c)}/\psi_J^{(c)} = \nabla\psi_J/\psi_J
+ (1-c)\nabla r/[c+(1-c)r]$, so the velocity is unchanged exactly when
Eq.~\eqref{eq:velocity_invariance_condition} holds on the common nonnodal domain, where $c+(1-c)r\neq0$. For $c>0$ write
$r = a + \im b$; the condition becomes
$c\operatorname{Im}\nabla r + (1-c)\operatorname{Im}(r^*\nabla r)=0$.
Requiring it for all $c>0$ gives $\operatorname{Im}\nabla r=0$, so $b$ is
spatially constant, and $\operatorname{Im}(r^*\nabla r) = -b\nabla a=0$.
If $b=0$, $r$ is real; otherwise $\nabla a=0$ and $r$ is constant.
Conversely both cases give invariance.
For complex $c$ with $\operatorname{Im}c\neq0$, suppose the velocity is
invariant for every such $c$. Fix $x$ with $\psi_J(x)\neq0$, so that $\psi_J^{(c)}(x)\neq0$ for $c$ near $1$, and let $A$ be
any component of $\nabla r(x)$. The condition reads
$\operatorname{Im}f(c)=0$ with
$f(c) = (1-c)A/[c+(1-c)r(x)]$. This $f$ is holomorphic where the
denominator is nonzero, which is a neighbourhood of $c=1$ (the
denominator equals $1$ there). On each open half-plane
$\{\operatorname{Im}c\gtrless0\}$ near $c=1$ the hypothesis gives
$\operatorname{Im}f=0$, so by the open mapping theorem for holomorphic functions \cite{conway1978complex} $f$ is constant
there. Letting $c\to1$, the numerator tends to $0$ and the denominator to
$1$, so the constant is $0$. Hence $(1-c)A=0$ for $c\neq1$, i.e.\ $A=0$.
Since $A$ and $x$ were arbitrary, $\nabla r=0$.
\end{enumerate}

\section{Numerical Implementation and Verification}
\label{app:numerics}
\label{sec:numerics}
\label{sec:num_method}
The numerical results require two distinct checks: that the computed guiding field solves the sourced Schr\"odinger equation, and that the transported particle ensemble satisfies the independent identities of the two-channel theory. We describe the algorithm, the scaled units and the accuracy limits in that order.

The analytical formulas throughout the paper retain $\hbar$, masses and frequencies. Numerical values instead use the family-specific scales declared in Sec.~\ref{sec:examples}, setting $\hbar=m=1$ in all examples; this choice does not render those constants dimensionless physically. For HO.1--HO.4 and 2B we also set $\omega=1$ and measure length, time and energy in $\ell_0=\sqrt{\hbar/m\omega}$, $t_0=1/\omega$ and $E_0=\hbar\omega$. The normalized-mode coefficient $j_0$ is quoted in $E_0$, $\tau=j_0t/\hbar$ is the weak-drive time and the HO.2 displacement $s=0.5$ is in $\ell_0$. The two-body coupling is $k/m\omega^2=0.5$ for $\tfrac12k(x_1-x_2)^2$.

The ring has no intrinsic frequency. With $L=2\pi$ we use $\ell_*=L/2\pi=1$, $t_*=m\ell_*^2/\hbar$ and $E_*=\hbar/t_*$. Its source coefficient obeys $j\sqrt L=\hbar(\Omega-\hbar q^2/2m)c_q$, with both terms in parentheses frequencies; in the scaled units this reads $j\sqrt{2\pi}=(\Omega-q^2/2)c_q$ (Sec.~\ref{sec:ring}). F$\infty$.2 uses its Gaussian width as $\ell_*$ and the corresponding $t_*$ and $E_*$. Because its spatial profile is dimensionless and $\int h_\varepsilon\,\mathrm dt=1$, its pulse coefficient has dimensions of energy times time per square root of length, and is expressed in $\hbar\ell_*^{-1/2}$, not in the oscillator's $E_0$. The labels F$\infty$.2, Ring, HO.1--HO.4 and 2B refer to the models in Secs.~\ref{sec:num_jump}, \ref{sec:ring}, \ref{sec:ho1} and \ref{sec:twobody}.

The source rate in Eq.~\eqref{eq:F_solution} must be sampled at the instantaneous particle position. We therefore integrate the trajectory and the logarithm of its ratio field together; advancing them separately would require interpolation of a near-nodal trajectory. The coupled characteristic system is

\begin{equation}
\frac{\mathrm{d}}{\mathrm{d}t}
\begin{pmatrix}
q \\[2pt] \ln F
\end{pmatrix}
=
\begin{pmatrix}
v^{\psi_J}(q,t) \\[4pt]
-\Gamma(q,t)
\end{pmatrix},
\qquad
\begin{pmatrix}
q(0) \\[2pt] \ln F(0)
\end{pmatrix}
=
\begin{pmatrix}
x_0 \\[2pt]
\ln f_0(x_0)
\end{pmatrix},
\label{eq:main_ode}
\end{equation}
The normalized ratio is then $\ln\tilde F=\ln F+\ln N_{\mathrm B}$. Adaptive step-size control acts on both state variables, so the integration has only one additional scalar beyond the $3N$ trajectory coordinates. Near a node the solver contracts its step; the finiteness of $\ln F$ through a passage remains conditional on integrability of $\Gamma$ along that trajectory. An alternative formulation using $\hat\Gamma$ would require the global average $\langle\Gamma\rangle_\rho$ at every step. Here $N_{\mathrm B}$ is evaluated independently from the field sector, so that extra global quadrature is unnecessary during trajectory integration.

For the driven models, $\psi_J$, its first two spatial derivatives and $J$ are evaluated from the Duhamel solution. F$\infty$.2 additionally requires a one-dimensional time integral, evaluated by 64-point Gauss--Legendre quadrature. The particle system~\eqref{eq:main_ode} then uses $v=\operatorname{Im}(\partial_x\psi_J/\psi_J)$ and $\Gamma=2\operatorname{Im}(J/\psi_J)$ in computational units.

We also propagate the flow Jacobian $M(t)=\partial q(t;x_0)/\partial x_0$, which measures deformation of a volume element carried by the trajectories. It satisfies $\dot M=(\nabla v)M$ and, in one dimension, $\mathrm d\ln|M|/\mathrm dt=\partial_xv$. Conservation of particle probability along the flow gives $P(q(t),t)=P_0(x_0)/|\det M|$. This provides an independent reconstruction of the transported density and of $D$, without integrating $\Gamma$.

Each sampled trajectory has an individual adaptive Dormand--Prince step.\footnote{The explicit Runge--Kutta pair uses orders five and four, whose difference supplies a local error estimate \cite{dormand1980family,hairer1993solving}.} This allows near-nodal paths to take small steps without slowing the entire ensemble. The relative tolerances are $10^{-9}$--$10^{-11}$ and the absolute tolerances $10^{-12}$--$10^{-13}$. The Born weight $N_{\mathrm B}=\int|\psi_J|^2\,\mathrm dx$ and the field-weighted rate $\langle\Gamma\rangle_\rho=N_{\mathrm B}^{-1}\int2\operatorname{Im}(J\psi_J^*)\,\mathrm dx$ are evaluated directly from the guiding field, never estimated from the particle paths. The latter would give an average over $P$, which equals the Born average only at structural equilibrium.

Ensemble averages use deterministic quadrature over initial positions, $\langle g\rangle_{P_0}\simeq\sum_kw_kP_0(x_0^{(k)})g(x_0^{(k)})$. We apply the trapezoidal rule on uniform grids; it converges exponentially for smooth periodic integrands \cite{trefethen2014trapezoidal}, but not for sharply sheared near-nodal data. No Monte Carlo sampling is involved.

The crucial internal check is that $D$ can be found in two different ways. Route~A integrates the rate along each characteristic, giving $D_A=\langle\ln F+\ln N_{\mathrm B}\rangle_{P_0}$. Route~B reconstructs $P=P_0/|\det M|$ from the flow and evaluates $D_B=\langle\ln P_0-\ln|\det M|-\ln\rho(q,t)+\ln N_{\mathrm B}\rangle_{P_0}$. It does not use $\Gamma$ and costs only one additional scalar per trajectory in one dimension. Unless stated otherwise, Sec.~\ref{sec:examples} quotes route~B.

We check the field equation and the ensemble dynamics separately (Table~\ref{tab:verify}). Field residuals are evaluated by finite differences and compared with an independent split-operator solver, whose error decreases quadratically under time-step halving \cite{feit1982solution,strang1968construction}. At the particle level, the identity $\ln F=\ln P_0-\ln|\det M|-\ln\rho(q)$ compares direct source-rate integration with flow geometry. We then compare Born-weight quadrature with its closed form and test $D(t)-D(0)=\int_0^t(\langle\Gamma\rangle_\rho-\langle\Gamma\rangle_P)\,\mathrm dt'$, whose terms are evaluated independently from the field, the trajectories and route~B.

A separate accuracy test compares the trajectories with SciPy's eighth-order Dormand--Prince integrator \cite{hairer1993solving,virtanen2020scipy}, giving maximum position deviations $\nv{scipyQ}$ at tight tolerance. Refining the solver reduces the position error from $\nv{ladQ0}$ to $\nv{ladQ2}$. Independently, $D(8)=\nv{DEightUni}$ on a uniform grid of $6401$ nodes compared with $\nv{DEightAdapt}$ using adaptive refinement. Most checks concern HO.1. For analytically specified fields, the residual and rate tests share the same formulas for $\psi_J$, except for the independent split-operator calculation; their agreement establishes consistency, not independent experimental verification of $\Gamma$ or $\rho$.

\begin{table}[t]
\centering
\small
\caption{Checks of the mathematics. Entries are maximum absolute deviations over all trajectories and output times (relative where noted). The HO.1 checks use $j_0/E_0=0.3$ and $\omega t\le 8$ with about $1.6\times10^{3}$ trajectories.}
\label{tab:verify}
\begin{tabular}{p{4.3cm} p{3.5cm} p{3.6cm}}
\toprule
Test & What it checks & Maximum deviation \\
\midrule
Residual $|\im\partial_t\psi_J-H\psi_J-J|$ by finite differences; the integrated balance of $N_{\mathrm B}$ follows from the same identity and agrees to $\nv{rateNB}$ &
field sector, all examples & $\le\nv{pdeMax}$ \\
Split-operator solver, error after halving $\Delta t$ twice &
field sector, independent solver &
HO.1 $\nv{spHOA}\to\nv{spHOC}$; ring $\nv{spRingA}\to\nv{spRingC}$;
2B $\nv{spTwoA}\to\nv{spTwoC}$ (second order); pulse
(Sec.~\ref{sec:num_jump}) $\nv{spPulseA}$ (round-off) \\
Identity $\ln F=\ln P_0-\ln|\det M|-\ln\rho(q)$; averaged over $P_0$ it gives the agreement of routes A and B for $D$, $\max_t|D_A-D_B|=\nv{DADB}$ &
ratio equation against flow geometry & $\nv{identDev}$ \\
$N_{\mathrm B}$ by field quadrature against $1+(j_0t/\hbar)^2$ &
amplitude diagnostic (relative) & $\nv{NBquad}$ \\
$D(t)-D(0)=\int_0^t(\langle\Gamma\rangle_\rho-\langle\Gamma\rangle_P)\,\mathrm{d}t'$ &
structural rate law & $\nv{rateD}$, limited by the time quadrature of the spiky $\langle\Gamma\rangle_P$ \\
\bottomrule
\end{tabular}
\end{table}

The accuracy of an individual trajectory must not be confused with the accuracy of an ensemble integral. Position integration reaches $10^{-10}$ or better, but the HO.1 initial-position quadrature for $D$ converges only to between $10^{-4}$ and $5\times10^{-4}$ across the drive scan when compared with a refined grid (Fig.~\ref{fig:ho1}d).

The difficulty comes from repeated near-nodal passages. Near a simple one-dimensional node, $\psi_J\simeq a(x-x^*)+b(t-t^*)$ with complex $a,b$. At a distance $\varepsilon$ the local velocity and rate can be $O(1/\varepsilon)$ over an interval of order $\varepsilon$, leaving a finite $O(1)$ contribution to $\int\Gamma\,\mathrm dt$ that depends on which side of the node the trajectory passes. The flow map $x_0\mapsto q(t)$ develops narrow regions of strong shear, where $\ln F$ and $\ln|\det M|$ vary much more rapidly than the Gaussian initial density. This scaling is heuristic for simple passages, not a proof of rate integrability or trajectory completeness; those remain conditional on Assumptions~\ref{ass:A}--\ref{ass:E}.

Every sampled trajectory in the oscillator calculations was integrated through its near-nodal passages without excluding paths from the numerical average, and its rate integral remained finite. Finitely many paths, however, cannot establish that exact nodal hits have zero ensemble measure or that the full flow is complete. The resulting ensemble integrand is nonsmooth, so the trapezoidal quadrature converges algebraically and oscillator $D$ values are quoted to at most three digits. The HO.1 weak-drive endpoint values $0.229$, $0.224$, $0.225$, $0.225$ correspond to $j_0/E_0=0.2,0.1,0.05,0.025$, at $\tau=200$ except $\tau=100$ for the last, and remain finite-horizon estimates. One-dimensional initial grids ranged from $4\times10^2$ nodes on the ring to $3.2\times10^3$ in the HO.1 scan, each checked against a second grid.

The ring gives a useful contrast because the computed field has no nodes. Its invariant $F\rho(v-u)$ is conserved to $\nv{ringInvU}$ in the untrapped case; in the trapped case the deviation is $\nv{ringInvT}$ ($\nv{ringInvTbig}$ where the invariant is not small). Near the attracting phase the invariant becomes small and numerical conditioning deteriorates, so machine-precision conservation is not asserted there.

The two-body calculation has a different source of quadrature error. The exact reduction of the initial ensemble average to the relative coordinate (Sec.~\ref{sec:twobody}) means that the two-dimensional grid reproduces a one-dimensional quadrature at the same spacing. Table~\ref{tab:twobody_refine} shows that its error in $D(10)$ is of order $10^{-3}$ and changes sign nonmonotonically with refinement; in contrast the one-dimensional reference reaches $0.1649$ to within $10^{-4}$ from about $800$ nodes onward. The maximum of $D$ over time is less sensitive and agrees to $10^{-5}$ across the grids.

The pointwise reduction is more accurate than its integral: two-dimensional relative-coordinate trajectories and $\ln F$ match their one-dimensional counterparts to $\nv{twoPtQ}$ and $\nv{twoPtF}$. We consequently quote $D(10)\simeq0.165$ from the converged one-dimensional reference and do not assign three-digit accuracy to the direct two-dimensional ensemble. More precise direct two-dimensional quadrature would require adaptation in the initial coordinate.

\begin{table}[t]
\caption{Two-body $D(t)$ at $\omega t=10$ against the number of nodes (spacing $h$ in units of $\ell_0$ in the relative coordinate) of the initial grid, and convergence of the one-dimensional reference. The first group uses the spacing of the direct two-dimensional integration; the one-dimensional values were obtained with the same nodes.}
\label{tab:twobody_refine}
\centering
\begin{tabular}{lcc}
\toprule
Initial grid & nodes in $y$ & $D(10)$ \\
\midrule
$h=0.225$ & 49 & 0.1745 \\
$h=0.15$ & 73 & 0.1625 \\
$h=0.10$ & 109 & 0.1663 \\
$h=0.075$ & 145 & 0.1664 \\
$h=0.05$ & 217 & 0.1615 \\
$h=0.025$ & 433 & 0.1638 \\
\midrule
1D reference & 401 & 0.1654 \\
 & 801 & 0.1648 \\
 & 1601 & 0.1649 \\
 & 3201 & 0.1649 \\
 & 6401 & 0.1648 \\
\bottomrule
\end{tabular}
\end{table}

\section{Operational Reconstruction of the Born Weight}
\label{app:measure}
The scale-covariance discussion leaves an operational question: can the Born weight be recovered from more than normalized position statistics? Within the ontology assumed here, it can in principle be reconstructed by combining the particle distribution $P$, the normalized guiding state and a calibrated exogenous source $J$. This is a conditional consistency construction, not a theory of measurement or a claim that such tomography is presently feasible.

The premise is that $\psi_J$ and $J$ are physically comparable fields, with $J$ calibrated in energy times wavefunction units and $|\psi_J|^2$ a density of the same physical kind as $P$. A tomographic reconstruction of a quantum state ordinarily returns a ray, not the absolute amplitude of the guiding field. Write $\psi_J(t)=A(t)\phi(t)$, with $\|\phi\|=1$ in an arbitrary phase convention and $|A|^2=N_{\mathrm B}$. The orthogonal projector $\Pi_\perp=1-|\phi\rangle\langle\phi|$ removes the component along that ray. Substituting in the ISE~\eqref{eq:ise} and projecting gives

\begin{equation}
A\,\nu=\Pi_\perp J,\qquad
\nu:=\im\hbar\,\Pi_\perp\partial_t\phi-\Pi_\perp H\phi,
\label{eq:meas_perp}
\end{equation}
The vector $\nu$ is determined by the ray and its time derivative: it is the part of the normalized-state evolution not supplied by the source-free Hamiltonian. Under $\phi\to e^{\im\alpha(t)}\phi$, $\nu$ changes by $e^{\im\alpha}$ while $A$ changes by $e^{-\im\alpha}$, leaving their product invariant. If $\Pi_\perp J\ne0$, the orthogonal source response identifies the complex amplitude algebraically,

\begin{equation}
A=\frac{\langle\nu,\Pi_\perp J\rangle}{\|\nu\|^2},\qquad N_{\mathrm B}=|A|^2,
\label{eq:meas_A}
\end{equation}
No integration constant is needed: with the source strength calibrated, the normalized state's response fixes $|A|^2$, not merely its rate. The nonzero orthogonal component $\Pi_\perp J$ is therefore the identifiability condition. When $\Pi_\perp J=0$, equivalently $J=w\phi$ with spatially constant $w$ (and $W=J/\psi_J$ uniform), the normalized projective dynamics carries no information on $A$: $\nu=0$ and Eq.~\eqref{eq:meas_A} cannot be used. This is the amplitude-only HO.2 case. It does not exclude the possibility of an independent initial calibration or other data constraining the amplitude. In this case $\Gamma$ is also spatially uniform, so the structural channel remains silent.

The projection along the ray supplies a complementary relation. With $\lambda=\langle\phi,H\phi\rangle-\im\hbar\langle\phi,\partial_t\phi\rangle$ real, it gives $\im\hbar\dot A=A\lambda+\langle\phi,J\rangle$ and hence

\begin{equation}
\dot N_{\mathrm B}=\frac{2}{\hbar}\operatorname{Im}\bigl(\bar A\langle\phi,J\rangle\bigr),
\label{eq:meas_par}
\end{equation}
This is precisely the Born-weight rate law in measured-ray variables. The reconstruction therefore has three pieces. The normalized state and calibrated source give $A$ through Eq.~\eqref{eq:meas_A}; the ensemble distribution and $|\phi|^2$ give $D=\int P\ln(P/|\phi|^2)\,\mathrm d\Omega$ independently of $N_{\mathrm B}$; and the source rate follows as $\Gamma=(2/\hbar)\operatorname{Im}(J/(A\phi))$.

The construction is overdetermined, which gives it testable consistency conditions. The vectors $\Pi_\perp J$ and $\nu$ must be complex-parallel; the recovered amplitude must obey the rate law~\eqref{eq:meas_par}; and the measured $\dot D$ must match $\langle\Gamma\rangle_\rho-\langle\Gamma\rangle_P$. A failure would challenge the combined field equation, prescribed source and guidance dynamics, not merely a normalization convention.

Example~HO.4 provides a direct test of the reconstruction in the same source-extended model. Its field $\psi_J=c\,e^{-\im\omega t/2}\varphi_0-\im(j_0t/\hbar)e^{-\im3\omega t/2}\varphi_1$ lies in the span of $\varphi_0$ and $\varphi_1$. In this two-dimensional subspace the orthogonal complement of the ray is one-dimensional, so $\Pi_\perp J$ and $\nu$ must be parallel. Evaluating Eq.~\eqref{eq:meas_A} from a finite-difference derivative of the model ray gives $|A|^2=c^2+(j_0t/\hbar)^2$ for $c=1,2,3$, to differentiation accuracy. Across four times for each $c$ and two phase conventions, the maximum absolute error in $N_{\mathrm B}$ is a few $10^{-9}$. This is a model calculation, not a laboratory validation.

The recovered initial weight $c^2$ distinguishes guiding fields of different amplitude at fixed $j_0$, as expected from the $j_0/c$ dependence of HO.4. The inference nevertheless relies on the ISE with known $H$ and $J$ and on a well-conditioned reconstruction of the ray and its derivative. It becomes ill-conditioned as $\|\Pi_\perp J\|$ becomes small, while many-body tomography is costly in a high-dimensional configuration space. The appendix does not establish the physical premise. It shows that, if that premise is adopted, the Born weight enters a definite set of mutually testable relations rather than being introduced solely by convention.

\section*{References}



\begin{thebibliography}{10}
\providecommand{\url}[1]{#1}
\csname url@samestyle\endcsname
\providecommand{\newblock}{\relax}
\providecommand{\bibinfo}[2]{#2}
\providecommand{\BIBentrySTDinterwordspacing}{\spaceskip=0pt\relax}
\providecommand{\BIBentryALTinterwordstretchfactor}{4}
\providecommand{\BIBentryALTinterwordspacing}{\spaceskip=\fontdimen2\font plus
\BIBentryALTinterwordstretchfactor\fontdimen3\font minus \fontdimen4\font\relax}
\providecommand{\BIBforeignlanguage}[2]{{%
\expandafter\ifx\csname l@#1\endcsname\relax
\typeout{** WARNING: IEEEtran.bst: No hyphenation pattern has been}%
\typeout{** loaded for the language `#1'. Using the pattern for}%
\typeout{** the default language instead.}%
\else
\language=\csname l@#1\endcsname
\fi
#2}}
\providecommand{\BIBdecl}{\relax}
\BIBdecl

\bibitem{Bacciagaluppi_Valantini2009}
G.~Bacciagaluppi and A.~Valentini, \emph{Quantum Theory at the Crossroads: Reconsidering the 1927 Solvay Conference}.\hskip 1em plus 0.5em minus 0.4em\relax Cambridge, UK: Cambridge University Press, 2009.

\bibitem{deBroglie1927_Solvay}
L.~de~Broglie, ``La nouvelle dynamique des quanta,'' in \emph{\'{E}lectrons et Photons: Rapports et Discussions du Cinqui\`eme Conseil de Physique, Solvay, 1927}.\hskip 1em plus 0.5em minus 0.4em\relax Paris: Gauthier-Villars, 1928, pp. 105--132, english translation in \cite{Bacciagaluppi_Valantini2009}.

\bibitem{Bohm1952Part_I}
D.~Bohm, ``A suggested interpretation of the quantum theory in terms of hidden variables -- {Part I},'' \emph{Phys. Rev.}, vol.~85, pp. 166--179, Jan 1952.

\bibitem{Bohm1952Part_II}
------, ``A suggested interpretation of the quantum theory in terms of hidden variables -- {Part II},'' \emph{Phys. Rev.}, vol.~85, pp. 180--193, Jan 1952.

\bibitem{bohm1993the_undivided_universe}
D.~Bohm and B.~J. Hiley, \emph{The undivided universe: an ontological interpretation of quantum theory}.\hskip 1em plus 0.5em minus 0.4em\relax London New York: Routledge, 1993.

\bibitem{durrr2009bohmian_mechanics}
D.~D{\"u}rr and S.~Teufel, \emph{Bohmian Mechanics: The Physics and Mathematics of Quantum Theory}.\hskip 1em plus 0.5em minus 0.4em\relax Berlin and Heidelberg: Springer, 2009.

\bibitem{maudlin2019quantum_philosophy}
T.~Maudlin, \emph{Philosophy of Physics: Quantum Theory}, ser. Princeton Foundations of Contemporary Philosophy.\hskip 1em plus 0.5em minus 0.4em\relax Princeton, NJ: Princeton University Press, 2019, eBook ISBN: 9780691190679.

\bibitem{bell2004speakable}
J.~S. Bell, \emph{Speakable and unspeakable in quantum mechanics: collected papers on quantum philosophy}.\hskip 1em plus 0.5em minus 0.4em\relax Cambridge New York: Cambridge University Press, 2004.

\bibitem{bohm1987quantum_implications}
B.~Hiley and F.~D. Peat, Eds., \emph{Quantum implications: essays in honour of David Bohm}.\hskip 1em plus 0.5em minus 0.4em\relax New York, NY: Published in the USA by Routledge \& Kegan Paul in association with Methuen, 1987.

\bibitem{Oldofredi2025Guiding_waves_in_quantum_mechanics}
A.~Oldofredi, Ed., \emph{Guiding Waves in Quantum Mechanics: One Hundred Years of de Broglie--Bohm Pilot-Wave Theory}.\hskip 1em plus 0.5em minus 0.4em\relax Oxford: Oxford University Press, 2025.

\bibitem{holland1995the_quantum_theory_of_motion}
P.~Holland, \emph{The quantum theory of motion: an account of the de {Broglie-Bohm} causal interpretation of quantum mechanics}.\hskip 1em plus 0.5em minus 0.4em\relax Cambridge, UK: Cambridge University Press, 1993.

\bibitem{Valentini1997Lorentz}
A.~Valentini, ``On {Galilean and Lorentz} invariance in pilot-wave dynamics,'' \emph{Physics Letters A}, vol. 228, no. 4-5, pp. 215--222, 1997.

\bibitem{Valentini2025The_trouble}
------, ``The trouble with pilot-wave theory: a critical evaluation,'' in \emph{Guiding Waves in Quantum Mechanics: One Hundred Years of de Broglie-Bohm Pilot-Wave Theory}, A.~Oldofredi, Ed.\hskip 1em plus 0.5em minus 0.4em\relax Oxford: Oxford University Press, May 2025, oxford Academic online edition published 22 May 2025.

\bibitem{Bohm1953}
D.~Bohm, ``Proof that probability density approaches {|$\psi$|${}^2$} in causal interpretation of the quantum theory,'' \emph{Physical Review}, vol.~89, no.~2, pp. 458--466, January 1953.

\bibitem{BohmVigier1954}
D.~Bohm and J.-P. Vigier, ``Model of the causal interpretation of quantum theory in terms of a fluid with irregular fluctuations,'' \emph{Physical Review}, vol.~96, no.~1, pp. 208--216, October 1954.

\bibitem{Valentini1991a}
A.~Valentini, ``Signal-locality, uncertainty, and the subquantum {H}-theorem. {I},'' \emph{Physics Letters A}, vol. 156, no. 1-2, pp. 5--11, June 1991.

\bibitem{Valentini1991b}
------, ``Signal-locality, uncertainty, and the subquantum {H}-theorem. {II},'' \emph{Physics Letters A}, vol. 158, no. 1-2, pp. 1--8, August 1991.

\bibitem{durr1992quantum_equilibrium}
D.~D{\"u}rr, S.~Goldstein, and N.~Zangh{\`i}, ``Quantum equilibrium and the origin of absolute uncertainty,'' \emph{Journal of Statistical Physics}, vol.~67, pp. 843--907, 1992.

\bibitem{Valentini2025beyond_the_quantum}
A.~Valentini, \emph{Beyond the Quantum: A Quest for the Origin and Hidden Meaning of Quantum Mechanics}.\hskip 1em plus 0.5em minus 0.4em\relax Oxford: Oxford University Press, September 2025, print ISBN: 9780198853749, Online ISBN: 9780191888175.

\bibitem{valentini2002subquantum_technology}
------, ``Subquantum information and computation,'' \emph{Pramana}, vol.~59, no.~2, pp. 269--277, 2002.

\bibitem{valentini2005dynamical}
A.~Valentini and H.~Westman, ``Dynamical origin of quantum probabilities,'' \emph{Proceedings of the Royal Society A}, vol. 461, pp. 253--272, 2005.

\bibitem{towler2011relaxation}
M.~D. Towler, N.~J. Russell, and A.~Valentini, ``Time scales for dynamical relaxation to the born rule,'' \emph{Proceedings of the Royal Society A}, vol. 468, pp. 990--1013, 2012.

\bibitem{valentini2008inflationary}
A.~Valentini, ``Inflationary cosmology as a probe of primordial quantum mechanics,'' \emph{Physical Review D}, vol.~82, p. 063513, 2010.

\bibitem{valentini2010inflationary}
------, ``Beyond the quantum,'' \emph{Physics World}, vol.~22, no.~11, pp. 32--37, 2009.

\bibitem{valentini2025dBB_Encyclopedia}
------, ``De {Broglie}-{Bohm} quantum mechanics,'' in \emph{Encyclopedia of Mathematical Physics}, 2nd~ed.\hskip 1em plus 0.5em minus 0.4em\relax Amsterdam: Elsevier, 2025, pp. 24--41.

\bibitem{mikki2026AdP}
\BIBentryALTinterwordspacing
S.~Mikki, ``de {Broglie--Bohm} dynamics with {Schr\"odinger} source fields: a framework for subquantum theory,'' \emph{Annalen der Physik}, 2026, in press. [Online]. Available: \url{https://arxiv.org/abs/2608.13894}
\BIBentrySTDinterwordspacing

\bibitem{prigogine2017non-equilibrium}
I.~Prigogine, \emph{Non-Equilibrium Statistical Mechanics}.\hskip 1em plus 0.5em minus 0.4em\relax Mineola: Dover Publications, 2017.

\bibitem{HallReginatto2016Ensembles}
M.~J.~W. Hall and M.~Reginatto, \emph{Ensembles on Configuration Space: Classical, Quantum, and Beyond}, 1st~ed., ser. Fundamental Theories of Physics.\hskip 1em plus 0.5em minus 0.4em\relax Cham: Springer, 2016, vol. 184, softcover reprint published in 2018 (ISBN 9783319816920).

\bibitem{Lindblad1976}
G.~Lindblad, ``On the generators of quantum dynamical semigroups,'' \emph{Communications in Mathematical Physics}, vol.~48, pp. 119--130, 1976.

\bibitem{GoriniKossakowskiSudarshan1976}
V.~Gorini, A.~Kossakowski, and E.~C.~G. Sudarshan, ``Completely positive dynamical semigroups of {N}-level systems,'' \emph{Journal of Mathematical Physics}, vol.~17, pp. 821--825, 1976.

\bibitem{deotto1998bohmian}
E.~Deotto and G.~C. Ghirardi, ``Bohmian mechanics revisited,'' \emph{Foundations of Physics}, vol.~28, no.~1, pp. 1--30, 1998.

\bibitem{Sontz2020Quantum_probability}
S.~B. Sontz, \emph{An Introductory Path to Quantum Theory: Using Mathematics to Understand the Ideas of Physics}.\hskip 1em plus 0.5em minus 0.4em\relax Cham: Springer, 2020.

\bibitem{cazenave2003semilinear}
T.~Cazenave, \emph{Semilinear {S}chr\"{o}dinger Equations}, ser. Courant Lecture Notes in Mathematics.\hskip 1em plus 0.5em minus 0.4em\relax New York and Providence, RI: Courant Institute of Mathematical Sciences and American Mathematical Society, 2003, vol.~10.

\bibitem{reed1980functional_1}
M.~Reed and B.~Simon, \emph{Methods of Modern Mathematical Physics}.\hskip 1em plus 0.5em minus 0.4em\relax New York: Academic Press, 1980, vol.~1, revised and enlarged edition.

\bibitem{berndl1995global_existence}
K.~Berndl, D.~D\"urr, S.~Goldstein, G.~Peruzzi, and N.~Zangh\`i, ``On the global existence of {Bohmian} mechanics,'' \emph{Commun. Math. Phys.}, vol. 173, no.~3, pp. 647--673, 1995.

\bibitem{pazy1983semigroups}
A.~Pazy, \emph{Semigroups of Linear Operators and Applications to Partial Differential Equations}, ser. Applied Mathematical Sciences.\hskip 1em plus 0.5em minus 0.4em\relax New York: Springer-Verlag, 1983, vol.~44.

\bibitem{mikki2026existence}
S.~Mikki, ``The subquantum fiber bundle: Skew-product dynamics and global existence of pilot-wave trajectories in source-extended de broglie--bohm theory,'' 2026, manuscript in preparation.

\bibitem{conway1978complex}
J.~B. Conway, \emph{Functions of One Complex Variable I}, 2nd~ed., ser. Graduate Texts in Mathematics.\hskip 1em plus 0.5em minus 0.4em\relax New York: Springer, 1978, vol.~11.

\bibitem{dormand1980family}
J.~R. Dormand and P.~J. Prince, ``A family of embedded {Runge--Kutta} formulae,'' \emph{J. Comput. Appl. Math.}, vol.~6, no.~1, pp. 19--26, 1980.

\bibitem{hairer1993solving}
E.~Hairer, S.~P. N{\o}rsett, and G.~Wanner, \emph{Solving Ordinary Differential Equations {I}: Nonstiff Problems}, 2nd~ed.\hskip 1em plus 0.5em minus 0.4em\relax Berlin: Springer, 1993.

\bibitem{trefethen2014trapezoidal}
L.~N. Trefethen and J.~A.~C. Weideman, ``The exponentially convergent trapezoidal rule,'' \emph{SIAM Rev.}, vol.~56, no.~3, pp. 385--458, 2014.

\bibitem{feit1982solution}
M.~D. Feit, J.~Fleck, J.~A., and A.~Steiger, ``Solution of the {Schr\"odinger} equation by a spectral method,'' \emph{J. Comput. Phys.}, vol.~47, no.~3, pp. 412--433, 1982.

\bibitem{strang1968construction}
G.~Strang, ``On the construction and comparison of difference schemes,'' \emph{SIAM J. Numer. Anal.}, vol.~5, no.~3, pp. 506--517, 1968.

\bibitem{virtanen2020scipy}
P.~Virtanen \emph{et~al.}, ``{SciPy} 1.0: fundamental algorithms for scientific computing in {Python},'' \emph{Nat. Methods}, vol.~17, pp. 261--272, 2020.

\end{thebibliography}
\end{document}